# Decentralization illusion in Decentralized Finance: Evidence from tokenized voting in MakerDAO polls.[1]

Xiaotong Sun[1]

Charalampos Stasinakis[2]

Georgios Sermpinis[3]

[1]University of Glasgow Business School, University of Glasgow, Gilbert Scott Building, Glasgow G12 8QQ, United Kingdom. Email: Xiaotong.Sun@glasgow.ac.uk.

[2]**Corresponding author**: University of Glasgow Business School, University of Glasgow, Gilbert Scott Building, Glasgow G12 8QQ, United Kingdom. Email: Charalampos.Stasinakis@glasgow.ac.uk,

[3]University of Glasgow Business School, University of Glasgow, Gilbert Scott Building, Glasgow G12 8QQ, United Kingdom. Email: Georgios.Sermpinis@glasgow.ac.uk.

April 2022

## Abstract

Decentralized Autonomous Organization (DAO) is very popular in Decentralized Finance (DeFi) applications as it provides a decentralized governance solution through blockchain. We analyze the governance characteristics in the Maker protocol, its stablecoin DAI and governance token Maker (MKR). To achieve that, we establish several measurements of centralized governance. Our empirical analysis investigates the effect of centralized governance over a series of factors related to MKR and DAI, such as financial, transaction, network and twitter sentiment indicators. Our results show that governance centralization influences both the Maker protocol, and the distribution of voting power matters. The main implication of this study is that centralized governance in MakerDAO very much exists, while DeFi investors face a trade-off between decentralization and performance of a DeFi protocol. This further contributes to the contemporary debate on whether DeFi can be truly decentralized.

---

[1] The online appendices can be accessed via: https://drive.google.com/file/d/1XAlMbpSLFMzsxYAl-JgZZqD71ICatfT_/view?usp=share_link

# 1. Introduction

As Bitcoin was proposed in 2008 (Nakamoto, 2008), blockchain has deeply changed financial markets. Various debates evolve around the potential democratization of financial services (Bollaert et al., 2021), blockchain competition and services improvement (Choi et al., 2020; Zhang et al., 2022;), and investment opportunities in new tokenized assets (Howell et al., 2020; Anyfantaki et al., 2021; Karim et al., 2022). It is widely accepted, though, that the main disruption lies in the disintermediation of financial institutions from their centralized role. The absence of centralized third parties, e.g., central banks, in the blockchain universe and circumventing traditional barriers of financial markets' participation are the major attributes of this market revolution. There is no or limited role for a central authority. Such decentralized frameworks theoretically allow all participants to be part of prominent decision making and spread risk among each other (Abdikerimova & Feng, 2022). Decentralization, therefore, is logically regarded as the core value proposition of blockchain (Harvey et al., 2021).

Decentralized Finance (DeFi) is blockchain-based financial applications which are designed to replicate most financial activities, e.g., lending and borrowing, in the traditional markets. Theoretically, governance in DeFi is decentralized since all members are decision makers. In traditional finance, governance is inevitably centralized, which can be the origin of several problems. The most intractable issue is probably the agency problem, where the owners and managers of an organization have different interests. Managers can pursue their own profits in the expense of the owners' benefits (Fama & Jensen, 1983). Therefore, the most challenging objective of governance is to align the interests between owners and managers. As discussed by Lee (2019), the decentralized nature of blockchain brings forward the idea of a 'token economy', where capital is better directed to those users actually contributing with content and services. Within the DeFi context, owners and managers are theoretically identical, which creates an opportunity to investigate the premises of this debate once again. Another crucial intersection between traditional finance and DeFi is the stablecoins and their links with the potential introduction of Central Bank Digital Currencies (CBDCs). Even if stablecoins are considered safer than other cryptocurrencies, central banks continue to scrutinize them. Are these tokens really needed to ensure DeFi liquidity or introducing CBDCs ensures financial stability from a stablecoin crash?

Stemming from this background, evaluating how efficient DeFi is a very crucial task. As Momtaz (2022) explain, one pathway to settling this debate is by examining the true decentralized nature of DeFi platforms. Despite the fact that the market size of DeFi exceeds 80 billion dollars (as of April 2022), the debate on whether decentralization is realistic or an illusion still stands (Aramonte et al., 2021). DeFi platforms showcase elements of centralization, usually in the form of 'governance tokens' and power concentration to large coin-holders. This phenomenon can lead to collusion among core decision makers during the governance process. It is obvious, then, that governance becomes a very critical dimension of the success of true decentralization in blockchain. Decentralized Autonomous Organization (DAO) is one popular solution to decentralized governance and decision making. In a DAO, all members are the owners of the organization and they have decision-making power to the development of it. Usually, the suggested changes will be written in the form of an Improvement Proposal (IP), which then is voted through an established poll, where all members can make public their option. DAO members state their option through governance tokens. Usually, these governance tokens are also tradable cryptocurrencies. The votes are weighted by the amount of governance token held by voters. In other words, governance in DAO is tokenized. Currently, DAO is one of the most common governance mechanisms adopted by DeFi (World Economic Forum, 2021).

The literature of governance in blockchain is voluminous. Usually, that debate evolves around the pros and cons of decentralization. Decentralization will result in lower speed of decision-making process, implying that the network becomes inefficient (Hsieh et al., 2017). Yermack (2017) argued that in practice, blockchain governance is not completely decentralized. In some extreme cases, the final decision is taken by only the core developers. For example, Bitcoin core developers once decided to lower the transaction fees without discussing with the related community (Gervais et al., 2014). Recently, Jiang et al. (2022) discuss blockchain governance by evaluating the trade-off between stability and efficiency through the prism of sensitivity to transaction fees. The authors suggest that the decentralized and audible nature of the blockchain transaction is attractive, but transaction fee movements have led to fork splits and endangered the system's stability. Their findings show that when users have balanced preferences between efficiency and stability, raising transaction fees reduces congestion in the platform. However, when it comes to DeFi platforms, market performance of DeFi platforms may be more crucial rather than efficiency. What would be the effect of powerful voters proposing and voting polls that serve their own interests? Tsoukalas and Falk (2020) and Carter and Jeng (2021) provide some insights on this question. Many blockchain-based platforms apply token-weighted voting mechanism relying on the premise that tokenized voting incentivizes users towards higher quality voting and improves system performance. The authors explain that this is not always correct as this voting mechanism discourages truthful votes and decreases the stability of the platform.

Though such papers provide both theoretical models and empirical evidence of governance centralization in blockchain, the literature surprisingly remains silent when it comes to centralization in DeFi. Positioning the centralized governance debate in the DeFi universe at the forefront of the literature is the main motivation of this paper. We focus on the Ethereum-based DeFi platform, Maker protocol, developed and managed by MakerDAO, as a case study. The rationale behind this choice is simple. MakerDAO is one of the most influential based DAOs. Since 2017, when the DeFi universe expanded exponentially, the Maker protocol has emerged as the leading lending protocol, which conceptually replicates the operation of a bank in cryptocurrency markets. In the Maker protocol, Maker (MKR) is the governance token. In terms of its value, one token equals to one vote in the proposed polls. Except from this tokenized value, Maker protocol issues the DAI, which is a stablecoin soft-pegged to US Dollar (MakerDAO, 2020). Currently, DAI is one of the most traded stablecoins, with daily number of transactions exceeding the ten thousand. Although the Maker protocol seems to be a big success of DAO, the way it is governed in practice has not been rigidly examined. To the best of our knowledge, this is the first paper that focuses on providing empirical evidence of centralized governance in DeFi.

To achieve that, we collect information for the Maker protocol governance, including all voters, their choice, and votes in Maker governance polls from $5^{th}$ August 2019 to $22^{nd}$ October 2021. Our empirical analysis follows two stages. The first stage is to examine governance polls by defining three novel measurements of centralized governance, namely voting participation, centralized voting power, and distribution of governance token. In the second stage, we investigate effect of centralized governance to the development of the Maker protocol. To achieve this, we expose MKR and DAI to several Maker-specific factors. These factors can be divided into several categories, including financial, transaction, network and Twitter sentiment indicators. Besides, we also investigate the ratios of collateral assets locked in Maker protocol. Such an empirical setup is consistent with similar studies in the field, such as Liu and Tsyvinski (2021). Beside well-investigated factors, e.g., network factors, we also consider transaction demand, which is a theoretical determinant of token price (Cong et al., 2020). Intuitively, transaction factors may relate to governance in DeFi. Finally, since users have to lock collateral assets before initiating loans from Maker, acceptable collaterals are a main issue discussed in Maker governance. If a risky on-chain asset is approved as collateral, theoretically, Maker protocol will be less safe.

The empirical framework brings forward some very interesting findings. By examining governance polls in the Maker protocol, we observe signals of centralized Maker governance. Comparing with the rapidly increasing number of users, voters are centralized in a small group of members, and the most dominant voters are heterogeneous in characteristics. The unevenly distributed voting power, as a preliminary signal of governance centralization, leads to our measurements of governance centralization in Maker protocol. By applying factor analysis, we find a complex nexus of effects of centralized governance around voting participation and distribution of voting power. Intuitively, more voters are a signal of larger voting participation, implying more decentralized governance. Voting participation can directly affect financial factors of MKR and DAI. For example, the DAI price decreases with more voters in governance polls. This suggests that the stablecoin can be affected from the participation in the polls and that decentralized governance could cause depegging. Centralized distribution of the governance token, i.e., MKR, can decrease trading volume of MKR and DAI, implying that centralization may bring more serious problems. After expanding our work on other indicators, we find centralized governance can lower the adoption of Maker Protocol. This is a serious issue, as the more centralized MakerDAO becomes, more users eventually discard its use. This paints a rather not optimistic picture not only for the long-term growth of Maker protocol but also for other DAO-governed DeFi platforms. Finally, voting power distribution appears to play a significant role towards ratios of collateral assets. Consequently, centralized voting power of large voters may change the proportion of main collateral assets (e.g., stablecoins) locked in the platform. Overall, all the above findings suggest that both the degree of MakerDAO's centralized governance and its performance pose a relevant trade-off among DeFi investors.

The remainder of our paper is organized as follows. Section 2 provides a summary of the governance process in MakerDAO. The dataset and the measurements of centralized governance in the Maker protocol are defined in section 3. The main empirical results are presented in section 4, while robustness checks are provided in section 5. Section 6 provides some concluding remarks. Finally, the appendix provides a description of the utilized factors, the ETH related analysis and robustness checks using an instrumental variable.

## 2. Governance in the Maker protocol
### 2.1 Decentralized Autonomous Organization (DAO), MakerDAO and Maker protocol

Decentralized Autonomous Organization (DAO) is a novel mechanism of organizational governance and decision-making. The DAO white paper is first given by Jentzsch (2016). Technically, DAO can be deployed on blockchain, and currently, most DAOs rely on Ethereum, which is a programmable blockchain. Ethereum's yellow paper is introduced by Wood (2014), and Ethereum users can write smart contracts in a Turing-complete programming language such as Solidity. By writing and executing smart contracts, users can actualize various interactions and functions, e.g., transactions on Ethereum. The programmable character enables the implementation of DAO. The core of DAO governance is based on standard smart contract code instead of human actors. In other words, DAO's governance is tokenized. In practice, DAO-based protocols usually have their own governance token and governance token holders can vote on changes to the protocols.

MakerDAO was created in 2014, and it has grown up to one of the most influential DAOs. The Maker protocol is multi-collateral lending system, and the protocol is governed by MakerDAO teams, including individuals and service providers (MakerDAO, 2020). Based on the functions of Maker protocol, the protocol is usually categorized as a Lending Protocol (LP), resembling banks in cryptocurrency markets. Simply, users can lend their tokens to LPs for economic incentives, e.g., interests. On the other hand, users can borrow tokens, and LPs usually require collateralization. The economic mechanism, mathematical models and the roles of LPs are well discussed by Bartoletti et al. (2020). Maker protocol issues DAI and the protocol is de facto a Multi-Collateral Dai (MCD) system. DAI is probably the most notable stablecoin, which is soft-pegged to the US dollar. Stablecoins, such as DAI, are cryptoassets designed to cope with the volatility of traditional cryptocurrencies and provide a bridge with fiat currencies (Wang et al., 2020). As MCD was launched in 2019, in Maker protocol every user can lock any supportive collateral such as ETH and a corresponding amount of DAI will be generated as debt.[2]

Except DAI, we are also interested in the MKR token. In practice, MKR plays two roles. On the one hand, MKR is the governance token of Maker protocol. MKR holders can vote on changes to the protocol. On the other hand, MKR contributes to the recapitalization of the system. MKR is created or destroyed through the automated auction mechanism of Maker protocol. When the debt of protocol is outstanding, MKR is created and sold for DAI. The protocol sells DAI for MKR and the surplus MKR is destroyed. At the inception of MakerDAO, one million MKR were issued. The protocol sets a threshold of minimum and maximum of MKR and the total circulated MKR always fluctuates between the thresholds.

**2.2 Maker governance structure and voting process**

An innovative selling point of the Maker protocol is decentralized governance. In the Maker protocol, governance can be divided into two parts: on-chain governance and off-chain governance. In on-chain governance, there are two types of votes, namely the Governance Polls and Executive Votes. Any MKR holders can vote using the Maker Protocol's on-chain governance system. Governance polls, which are about non-technical changes, measure the sentiment of MKR holders. Executive votes "execute" technical changes to the protocol. The voting results are documented on blockchain. Off-chain governance is mainly about informal discussion, e.g., discussion on the MakerDAO forum. Both MKR holders and the larger community can express their opinions. Voting power is weighted by the amount of MKR that a voter owns and represents, making the voting mechanism a token-weighted one (Tsoukalas & Falk, 2020). One MKR equals to one vote and the option with the largest votes wins. In the Maker protocol, Maker IPs are structured and formalized for a voting event, and key issues and changes to the system are rigidly defined in Maker IPs. Usually, the Maker Foundation will draft the initial Maker IPs, and any community members can propose competing IPs. Then, the final decisions will be made by MKR voters through the current Maker governance process. The above information is illustrated in figure 1.

MKR holders can be voters and directly choose their options on Maker Governance Portal. On the other hand, they can choose a Vote Delegate to be their representatives. As a result, delegates gain voting power from MKR holders, and these MKR holders can indirectly vote. The voting results are weighted by the amount of MKR voted for a proposal. Noticeably, Vote Delegates were not introduced in the very beginning. On July 30[th], the guidance to Vote Delegates was first introduced in the Maker forum, while on 10[th] November 2021, there are 16 Delegates and 65989.65 MKR are delegated. Currently, there are three types of voting in Maker governance, i.e., Forum Signal Threads, Governance Polls and Executive Votes. These are summarized on table 1.

---
[2] As of April 2022, DAI is in fifth in market capitalization among all stable coins (over nine billion US dollars). For more details regarding of DAI debt in Maker protocol, we refer the reader to Qin et al. (2021).

Figure 1. Governance in Maker Protocol

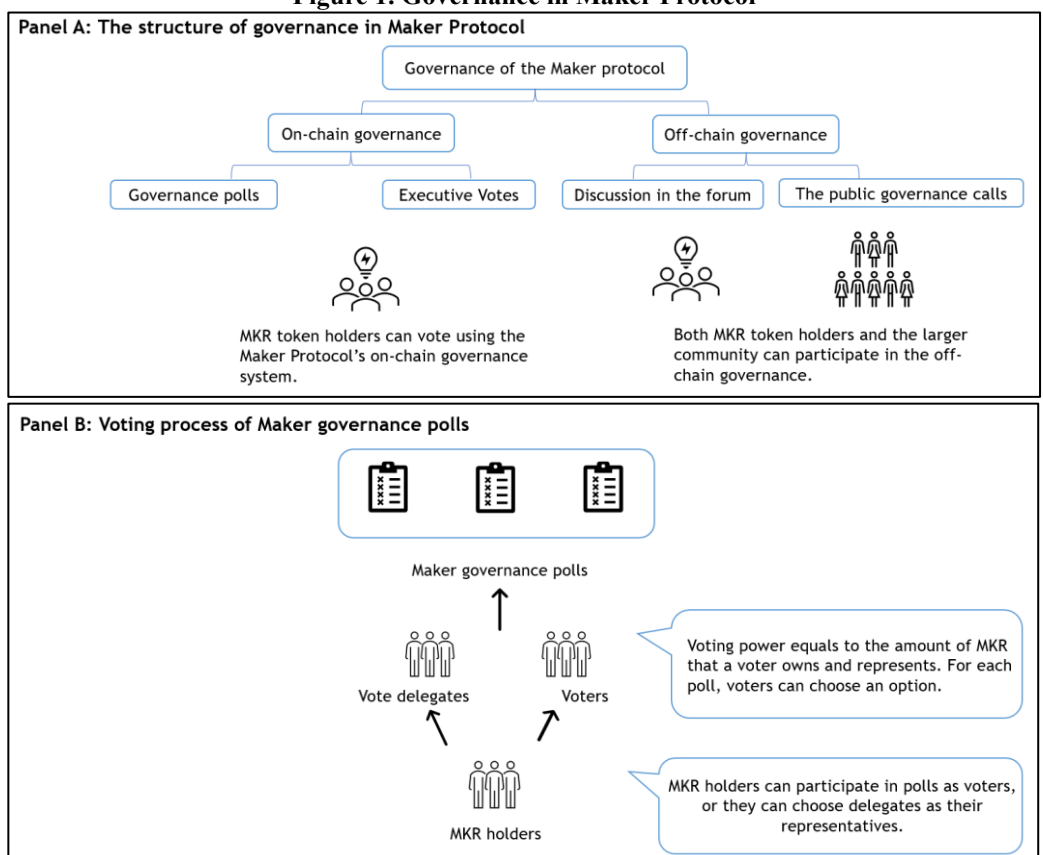

Note: Panel A shows the governance structure of Maker Protocol. It is divided into two parts, the on-chain governance and off-chain governance. Panel B illustrates the voting process of Maker governance polls. MKR holders can participate in polls as voters, or they can choose delegates as their representatives.

Table 1. Types of votes in Maker protocol

| Type of votes | Functions |
| --- | --- |
| **Forum signal threads** | (1) Determine consensus that something needs to be done in response to a perceived issue, (2) determine consensus for a concrete action to be taken in response to a perceived issue. |
| **Governance polls** | (1)Determine governance and DAO processes outside the technical layer of the Maker Protocol, (2) form consensus on important community goals and targets, (3) measure sentiment on potential Executive Vote proposals, (4) ratify governance proposals originating from the MakerDAO forum signal threads, (5) determine which values certain system parameters should be set to before those values are then confirmed in an executive vote, (6) ratify risk parameters for new collateral types as presented by Risk Teams. |
| **Executive votes** | (1) Add or remove collateral types, add or remove Vault types, adjust global system parameters, adjust Vault-specific parameters, (2) replace modular smart contracts. |

Note: This table describes three types of votes in Maker protocol. Forum signal threads are a part of off-chain governance, and all community can participate in the discussion on the Maker forum. Governance polls and executive votes are on-chain.

Table 1 highlights the functions of the three types of votes. Forum Signal Threads are the least consequential, and the threads are a part of off-chain governance. Governance Polls and Executive Votes occur on-chain, and they can be accessed through the Maker Foundation's Voting portal. Simply, Governance Polls determine processes outside the technical layer, while Executive Votes are about technical changes to the protocol. Executive Votes use 'Continuous Approval Voting' model to make the system more secure. The voting models means that new proposals need to surpass the voting weight of the last successful proposal (MakerDAO, 2021).

## 2.3 Governance centralization

Centralization in the governance layer of blockchain has caused attention of academic audience, and the discussion mainly focuses on two problems, namely owner control and improvement protocol (Sai et al., 2021). In Bitcoin, Gervais et al. (2014) argue that the author Satoshi Nakamoto may accumulate significant Bitcoin since Nakamoto participated in activities at the early stage of Bitcoin blockchain. Similar evidence of owner control exists in Ethereum as well (Bai et al., 2020). The large proportion of wealth may result in economic manipulation in blockchain. As for improvement protocol, this problem derives from the process of moderation in blockchain. Usually, blockchain and DeFi adopt improvement proposal system in decision making process. By analyzing the authors and contributors of improvement proposals, Azouvi et al. (2018) found that core developers are the main contributors to the development of Bitcoin and Ethereum. In other words, these core developers have more decision-making power in the decision-making process.

Beside the two problems described above, DAO, as a new organizational form to automate governance, may bring both opportunities and challenges. Benefiting from blockchain technology, the ownership is more transparent and voting can be more accurate (Yermack, 2017). On the other hand, centralization seems to be still inevitable in DAO. Maker Governance Polls do not attract much participation and MKR held by the dominant voters may theoretically lead to collusion in some poll. However, in order to examine this, we need information from many polls and this is not an easy task. As shown in the next section, the quest to obtain information from more polls is achieved in this study.

## 3. Data collection and identification

### 3.1 Data Collection

In Maker Governance Portal, the details of governance polls, e.g., the number of voters and results, are publicly available. To get the voters' addresses, we query the voting history from MCD Voting Tracker. We investigate governance polls from Poll 16 (deployed on 5$^{th}$ August 2019) to Poll 663 (deployed on 22$^{nd}$ October 2021). The dates when the polls are set are used to identify when the transactions are added to Ethereum blockchain. Though the dates might be earlier than the start dates when voters can choose options, the contents of polls are already publicly accessible once the polls are sent to the blockchain. Poll 16 is the first governance poll that MKR holders can participate in. Some polls failed[3], so they are not documented in the portal. Hence, the dataset consists of a total of 638 successful governance polls, and the voters' public names can be found by searching for their addresses on Etherscan.io and Maker Governance Portal. To study the effects of centralization in Maker governance, we consider two influential crypto assets issued by Maker protocol, namely MKR and DAI.[4]

### 3.2 Measurements of centralized voting power in Maker protocol

This section introduces the novel measurements of centralized voting power in Maker protocol, and the measurements can fall into three categories, i.e., voting participation, centralized voting power, and distribution of governance token. For each measurement in the first two categories, we first calculate the value at the poll level. Then, daily measurements can be generated. The distribution of governance token can divulgate more information of centralized power of certain Maker protocol users, e.g., MakerDAO delegates and large MKR holders.

---
[3] Poll 28, 39, 47, 69, 78, 183, 282, 284, 286, and 500 failed.
[4] The Maker portal is available at: https://vote.makerdao.com. The voting tracker is available at: https://beta.mcdgov.info. The DAI and MKR statistics can be found at: https://www.intotheblock.com/.

### 3.2.1 Voting participation

To proxy voting participation, we use two measurements. One is the total votes of Maker governance polls on a given date. The other is the number of total voters on a given date. Intuitively, when these two measurements are higher, there are more voters and votes in governance polls. Assuming that there are $n$ polls and $m$ voters on a date $d$, we have:

$$Total\ votes_d = \sum_{i=1}^{n} Total\ votes_{i,d} \quad (1) \quad \text{and} \quad Voters_d = \sum_{i=1}^{m} Voters_{i,d} \quad (2)$$

### 3.2.2 Centralized voting power

In order to capture centralized voting power, we utilize two measures. The first is the Gini coefficient which is traditionally used to measure inequality (Dorfman, 1979). Assuming that there are $l$ voters in a governance poll we have:

$$Gini = \frac{\sum_{i=1}^{l} \sum_{j=1}^{l} |votes_i - votes_j|}{2l^2 \overline{votes}} \quad (3)$$

Where $votes_i$ is the votes of voter $i$, and $\overline{votes}$ is the average votes in a governance poll.

After computing the Gini coefficient for each poll, we can calculate a daily measurement by calculating the average. Assuming that there are $n$ polls on a date $d$, we will calculate daily Gini coefficient, i.e., $Gini_d$, via maximum likelihood estimation (see, Taleb (2015)). The number of voters participated in governance polls can be very different, and bias and error can happen if we choose arithmetic means to measure daily Gini coefficient. Maximum likelihood estimation, as an indirect method, can have considerably lower error rate, especially when the sample sizes (in this case, the voters in a governance poll) vary.

The second proxy for centralized voting power is the largest voter's power in Maker governance polls. This can be approximated by the $Largest\ voting\ share$. In that way, we can reflect also on the relative voting power of the largest voter. For each poll, $Order$ refers to the voting order of the largest voter. When $Order$ is smaller, the largest voter will choose their option earlier. Assuming that there are $n$ polls on a date $d$, we have:

$$Largest\ voting\ share_d = \frac{\sum_{i=1}^{n} Largest\ voting\ share_{i,d}}{n} \quad (4)$$

Finally, the voting sequence can actually play a role as this can be documented in several voting systems (see amongst others, Börgers (2010), Brams (2008), and Brams and Fishburn (2002)). Simply, voters have different strategies, and their preference of voting order will vary with their goals. For each poll, we define a variable $order$ to measure the decision speed of the largest voter. Here, the order of the largest voter refers to the order in the whole history. It is to say, the largest voters may change their choice later. Assuming that there are $k$ records in the voting history of a governance poll $i$ on a date $d$, we have:

$$Order_{i,d} = \frac{voting\ order\ of\ the\ largest\ voter}{k} \quad (5)$$

Assuming that there are $n$ polls on a date $d$, we have

$$Order_d = \frac{\sum_{i=1}^{n} Order_{i,d}}{n} \quad (6)$$

When $order$ is smaller, the largest voter chooses an option earlier than other voters. All voters can see the existing choices on Maker Governance Portal.

### 3.2.3 Distribution of governance token

Beside the measurements stemming from voting behavior in Maker governance polls, we also consider distribution of governance token in MakerDAO (i.e., MKR), which reveals more information of centralized decision-makers. Nadler and Schär (2020) shown that token distribution is usually centralized in DeFi, and we suspect such centralization also exists in MakerDAO. First, we calculate the balance of MKR controlled by MakerDAO delegates, which equals to the sum of MKR owned by delegates and represented by delegates.[5] Assuming that there are $l$ delegates, the MKR balance controlled by these delegates on a date $d$ is:

$$delegate_d = \sum_{i=1}^{l} delegate_{i,d} \quad (7)$$

where $delegate_{i,d}$ is the MKR balance controlled by delegate $i$ on date $d$. A higher $delegate$ means more voting power is controlled by MakerDAO delegates, implying governance centralized caused by these influential MakerDAO users.

Then, we compute the proportion of MKR controlled by large MKR holders. Here, we consider three categories of MKR holders, including holders with a balance more than 10,000 MKR, holders with a balance between 10,000 and 100,000 MKR, and holders with a balance more than 100,000 MKR. On a date $d$, we assume that there are $x, y, z$ MKR holders in the three categories, respectively. Then, three measurements can be calculated to reflect on the centralized distribution of MKR:

$$>10k_d = \frac{\sum_{i=1}^{x} >10k_{i,d}}{total\ circulated\ MKR_d} \quad (8)$$

$$10k-100k_d = \frac{\sum_{i=1}^{y} 10k-100k_{i,d}}{total\ circulated\ MKR_d} \quad (9)$$

$$>100k_d = \frac{\sum_{i=1}^{z} >10k_{i,d}}{total\ circulated\ MKR_d} \quad (10)$$

where $total\ circulated\ MKR_d$ is the amount of MKR circulating on Ethereum blockchain on date $d$.

When the three measurements above are higher, more voting power is controlled by large MKR holders. Though governance polls are not deployed on a daily basis, the centralized distribution of MKR is a signal of governance centralization, indicating that voting in MakerDAO is dominated by large MKR holders.

---

[5] We query the voting power of MakerDAO delegates from dune.xyz. The query can be viewed on https://dune.com/queries/2023685

## 4. Empirical results

This section summarizes the empirical results of this study. The first sub-section presents the descriptive statistics of both polls and voters. Then, the centralization in Maker governance polls is described by the calculations of the measurements of centralized governance defined in the previous sections. The second sub-section summarizes the factor analysis we perform to investigate the effects of centralized governance on the Maker protocol (MKR, DAI and locked collateral assets).

### 4.1 First stage: Governance polls in the Maker protocol

The collected information from the 638 governance polls is crucial for understanding centralization in Maker protocol. Table 2 presents the descriptive statistics of these polls.

Table 2. Descriptive statistics of Maker governance polls

|  | Total votes | Total voters | Breakdown votes | Breakdown ratio | Breakdown voters | Votes of the largest voter | Vote share of the largest voter |
|---|---|---|---|---|---|---|---|
| **Mean** | 36096.52 | 24.59 | 31529.94 | 88.78% | 18.03 | 16941.61 | 52.66% |
| **Median** | 33097.15 | 23.00 | 28625.80 | 98.24% | 16.00 | 17063.93 | 48.35% |
| **Maximum** | 131555.35 | 146 | 108694.07 | 100.00% | 142 | 39403.85 | 98.51% |
| **Minimum** | 259.74 | 5 | 232.80 | 13.04% | 1 | 203.27 | 20.28% |
| **Std** | 22383.18 | 12.76 | 19998.47 | 16.67% | 11.63 | 8452.45 | 18.02% |

Note: This table reports the descriptive statistics of Maker governance polls. Breakdown votes refers to the votes of the winning option, and breakdown ratio is breakdown votes divided by total votes. Besides, breakdown voters are the number of voters who choose the winning option.

The votes are calculated in MKR tokens and for each governance poll, breakdown ratio is the proportion of breakdown votes to total votes. Beside the votes and vote share of the largest voter, the order of voting is considered. We also present the daily number of governance polls and voters in figure 2.

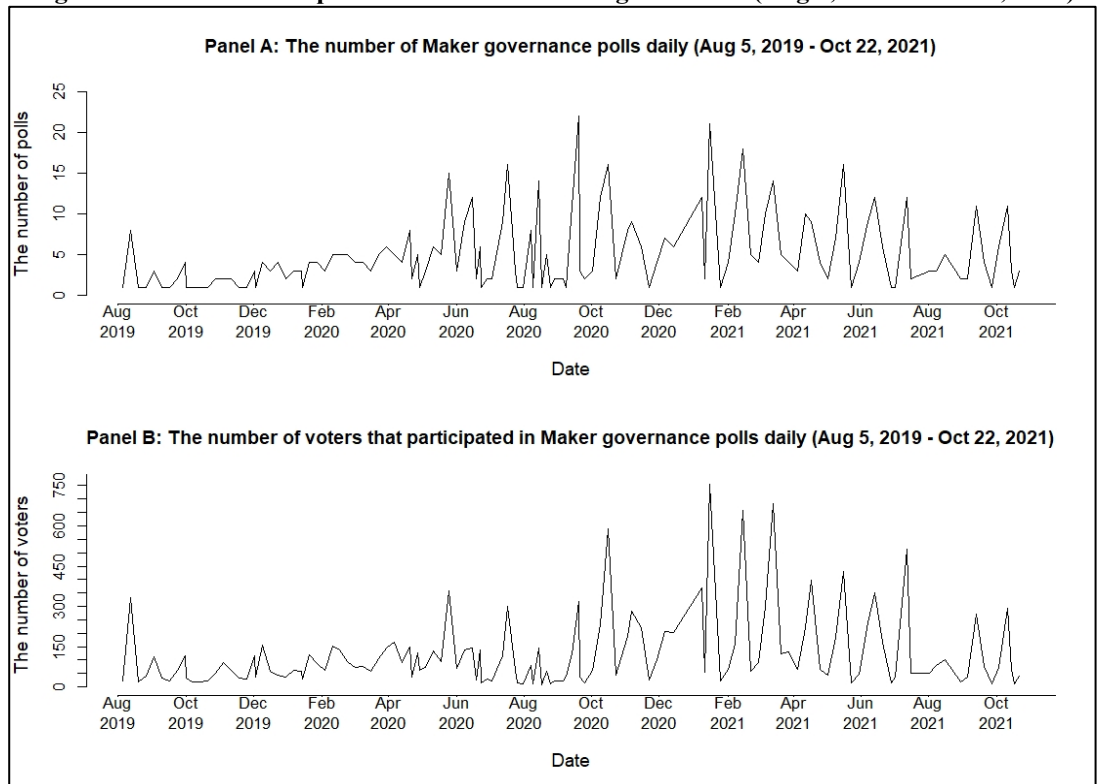

Figure 2. The number of polls and voters in Maker governance (Aug 5, 2019 – Oct 22, 2021)

Note: This figure presents the daily number of polls and voters in Maker governance. Panel A shows the daily number of Maker governance polls, while Panel B presents the number of voters daily in Maker governance polls.

From the figures we can easily see that within a day, the number of deployed polls is usually less than 25. Usually, no more than 700 voters will express their choices on the same day. For some polls, no more than ten voters will participate in decision making process. The finding implies that not all polls have large voting participation. Compared to the rapidly growth of Maker users, voters are a small group. Our analysis extracts a total of 1250 unique voters in our dataset. For each voter, the number of polls that they participate in can be surprisingly different. To show case this we present the following descriptive statistics in table 3.

By examining the total votes and the highest votes that a voter has in a single poll, it is implied that the voting power is not equally distributed across voters. This could be an early sign of voting centralization. However, to make this claim clearer, we need to delve deeper in the composition of the voters and their characteristics. To that end, we identify the voters whose identity is publicly available, the top ten voters that participate in most polls, the top ten voters that largest total votes and the top ten voters that have the largest single vote. This information is summarized in the online Appendix OA.1.

**Table 3. Descriptive statistics of voters of Maker governance polls**

|  | Involved polls | Total votes | First poll | The highest votes |
|---|---|---|---|---|
| **Mean** | 12.55 | 18422.58 | 278.66 | 665.39 |
| **Median** | 2.00 | 1.42 | 248.00 | 1.00 |
| **Maximum** | 514 | 4170786.51 | 660 | 39403.85 |
| **Minimum** | 1 | 0.00 | 16 | 0.00 |
| **Std** | 42.46 | 164269.26 | 194.65 | 3372.75 |

Note: This table reports the descriptive statistics of voters of Maker governance polls. For each voter, we calculate the number of polls that they participate in, their total votes, and the highest votes in a single poll. Here, votes are calculated in Maker (MKR), which is the governance token of the Maker Protocol. Besides, the first poll that a voter participated in is presented. A lower number means that the voter started to participate in Maker governance polls earlier.

We have some interesting findings towards identifying centralized governance in the Maker protocol. Except *a16z*[6], the known voters are delegates in Maker governance and their identity is publicly available on Maker governance portal, and more details of these known voters in governance polls are given as well. The mechanism of voting delegates was introduced in July 2021, therefore, most delegates participated in their poll in August 2021. Noticeably, the total votes and the highest votes in a single poll are different among these known voters. Field Technologies, Inc. is the known voter that has the largest total votes (as of November 1st, 2021). In terms of voters that are participating in most polls, it is obvious that their characteristics are different, while none of them has a public name, i.e., their identity is unknown. Voters with the largest total votes are again heterogeneous in characteristics, while only two from the top ten are found to be delegates (Field Technologies, Inc. and a shadow delegate). When accounting the voters with the largest single votes, we find again a different composition. We identify delegates such as Field Technologies, Inc., Flip Flop Flap Delegate LLC, a shadow delegate and *a16z* being dominant, while the remaining voters appear with unknown identities. Taking a wholistic look at these findings, we notice that some voters may both participate in many polls and have large total votes, namely voters with the addresses *0x4f...3f30* and *0x6a...ab40*. In other cases, some voters might not participate in many polls, but when they do, they have significantly large votes in certain polls. For example, *a16z* only votes for three polls, but their single votes are more than 30,000. These characteristics of the dominant voters suggested that on-chain developments on the protocol are driven by dominant voters and that decentralization does not seem to hold. Voting power appears to be distributed unevenly across different known or unknown small groups of voters, especially when total votes and large votes in a single poll are considered.

---
[6] It is easy to establish by searching for other voters' addresses on *Etherscan.io* that *a16z* represents the venture capital firm Andreessen Horowitz, which is the most influential venture capital in DeFi markets.

In order to further show this, we focus now on the notion of centralized voting power in the Maker polls. We compare the largest vote for each poll with the total votes and we find that the largest voter can account for a significant share of the total votes in most polls. Practically, the largest voter is the pivotal figure in implementing protocol changes, as they tend to account for around one third of the voting share. In terms of the known voters (namely delegates and *a16z*), the trend is similar. These known voters are identified after the delegate regulation change in Maker protocol (after Poll 600) and their dominant power is evident. However, it is hard to say if they were able to play an important role in previous polls. All this information is illustrated in the following figure.

**Figure 3. Total votes, votes of the largest voter and votes of the known voters in Maker polls (Poll #16 – Poll #663)**

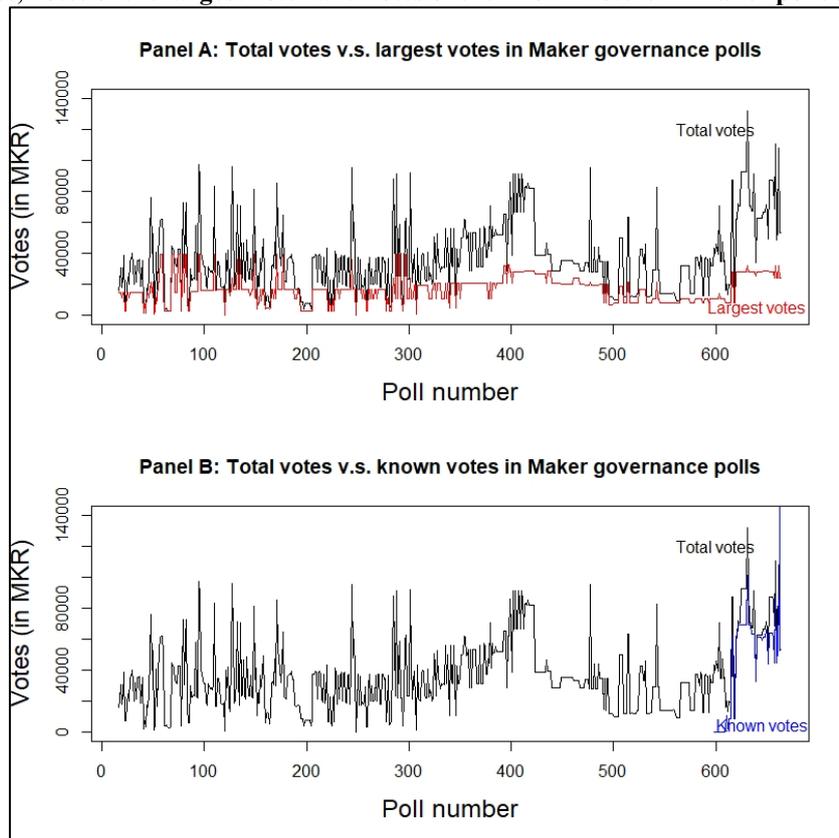

Note: Panel A presents the total votes and votes of largest voters in Maker governance polls (Poll #16 – Poll #663). In most polls, the largest voter can take account for a significant proportion of voting power. Panel B shows the votes from the known voters in Maker governance polls (Poll #16 – Poll #663). The known voters include voting delegates and a16z and show strong voting power after Poll #600.

To support the above, we also illustrate the total votes over the breakdown votes and their respective voters, the breakdown ratio, the voting share of the largest voter and average voting share of the largest voter daily. The results show that winning polls are driven by most votes, largest voters contribute significant votes to the winning options, while largest voters consistently concentrate at least 30% of the average daily voting share. These voting patterns are presented in the online Appendix OA.1. The key message remains that centralized voting power exists.

Although the above could be though descriptive information extracted from our unique dataset, we take further steps to quantitatively establish centralized governance on the Maker protocol. Firstly, we measure the centralized voting power in Maker governance at a poll level and across days by utilizing the Gini coefficient estimations. The results are summarized in the table and figure that follows.

**Table 4. Gini coefficient in Maker governance polls**

|  | Poll-level | Daily |
|---|---|---|
| **Mean** | 84.38% | 18.57% |
| **Median** | 85.54% | 0.00% |
| **Maximum** | 98.05% | 94.04% |
| **Minimum** | 57.56% | 0.00% |
| **Std** | 0.06 | 0.35 |

Note: This table reports the Gini coefficient in Maker governance polls. In the first column, we calculate the Gini coefficient for each poll. In the second column, we first integrate a voter's votes within a day, then we compute the daily Gini coefficient via a maximum likelihood estimation.

**Figure 4. Gini coefficient**

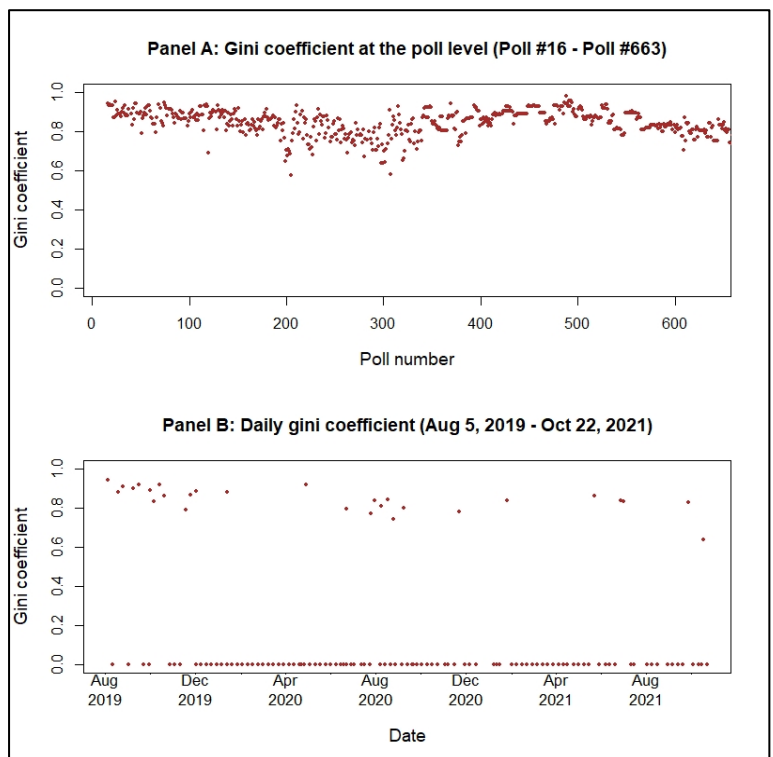

Note: This figure shows the Gini coefficient in Maker governance. Panel A reports the Gini coefficient at the poll level (Poll #16 – Poll #663). Panel B reports the Gini coefficient daily in Maker governance polls (Aug 5, 2019 – Oct 22, 2021).

At a poll level, we find that the Gini coefficient is always more than 50% and exhibits a maximum of 98.05%. Given that the Gini coefficient estimation is higher than 0.60 for most of the polls, highly centralized voting power in the Maker governance is established. We also calculate and illustrate the daily Gini coefficient. The expected daily average Gini coefficient should be around zero, if no centralized voting occurs. However, we observe that there are days that the value is higher than 0.75, implying again strong centralization of voting power in particular days within our period under study. We further highlight the evidence of vote centralization by estimating the Lorenz curve of the cumulative total votes for particular polls. The results support the above findings and are presented also in the online Appendix OA.1.

Besides, we also illustrate voting power of large MKR holders and MakerDAO delegates. For MKR holders whose MKR balance is between 10,000 and 100,000 (hereafter, major holders), their voting power is around 25%. For the MKR holders with more than 100,000 MKR (hereafter, whales), their voting power accounts for a significant proportion, though their voting power decreased since December 2020. The amount of MKR controlled by MakerDAO delegates should not be ignored, given that the spikes of their MKR balances are closed to 150,000 MKR.

**Figure 5. Voting power of large MKR holders and MakerDAO delegates**

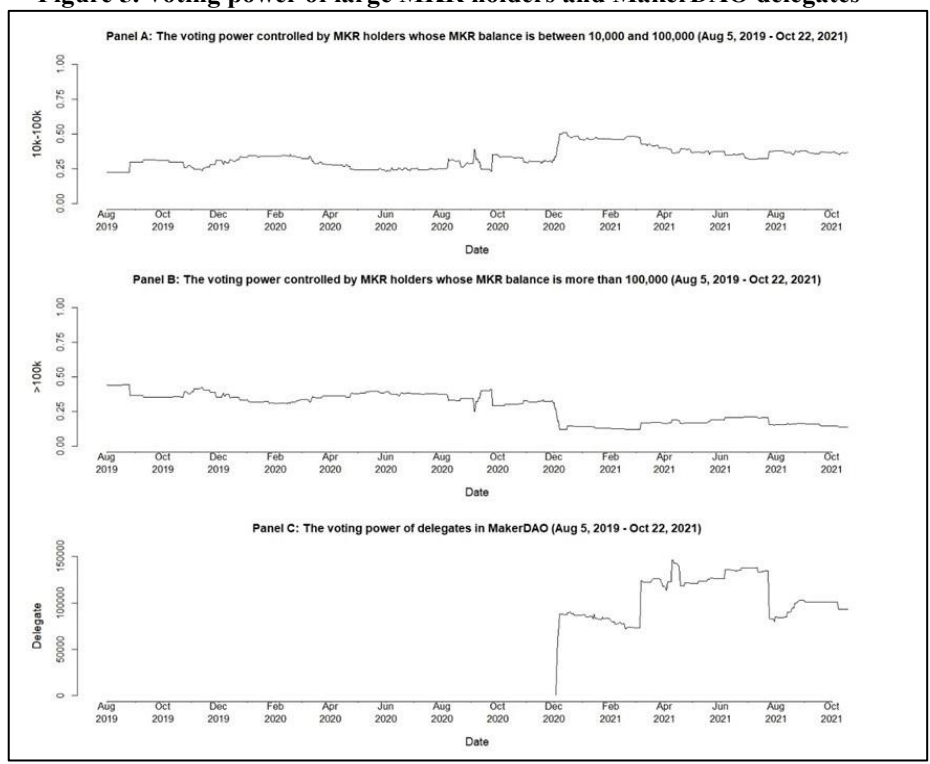

Note: This figure illustrates the proportion of MKR controlled by large MKR holders and MKR balance controlled by MakerDAO delegates (Aug 5, 2019 – Oct 22, 2021). In Panel A, we calculate the proportion of MKR controlled by holders whose MKR balance is between 10,000 and 100,000. In Panel B, we focus on MKR holders whose MKR balance is more than 100,000. Panel C shows the MKR balances controlled by MakerDAO delegates.

Finally, the other measurements of governance centralization are established based on the definitions given in Section 3. Their descriptive statistics are provided in the following table.

**Table 5. Measurements of governance centralization in Maker**

|  | Voters | TotalVotes | LargestShare | Order | 10k-100k | >100k | >10k | Delegate |
|---|---|---|---|---|---|---|---|---|
| **Mean** | 123.53 | 181335.26 | 0.54 | 0.41 | 0.33 | 0.28 | 0.61 | 105743.35 |
| **Median** | 68.00 | 114304.83 | 0.51 | 0.39 | 0.33 | 0.32 | 0.63 | 101145.62 |
| **Maximum** | 756.00 | 1251962.15 | 0.96 | 0.91 | 0.51 | 0.44 | 0.67 | 146462.93 |
| **Minimum** | 7.00 | 259.74 | 0.27 | 0.00 | 0.22 | 0.12 | 0.49 | 1151.71 |
| **Std** | 141.27 | 209387.53 | 0.16 | 0.20 | 0.07 | 0.10 | 0.05 | 23047.70 |
| **N of obs.** | 127 | 127 | 127 | 127 | 810 | 810 | 810 | 320 |

Note: This table presents the descriptive statistics of measurement of governance centralization in Maker. The first four columns report measurement of governance centralization in Maker governance polls. We first calculate these measurements for each poll and then convert it to daily level measurements. For example, we first calculate the number of voters for every poll, then we sum up to the get the number of daily voters. The last three measurements reflect on the influence of large MKR holders and delegates.

To simplify the factor analysis, we implement *Principal Component Analysis (PCA)*. Simply, higher explained variance ratios means the higher importance of measurements. In the following sections, we will estimate regressions using the four measurements, including $Gini, Voters, 100k - 100k, > 100k$. Besides, $delegate$ will be investigated[7].

---
[7] The regression results for other measurements can be provided upon request.

Table 6. Principal Component Analysis (PCA) of measurements

| Panel A: Measurements related to governance polls | | | |
|---|---|---|---|
| | Explained Variance Ratio | Explained Variance | N of obs. |
| **Gini** | **10.00e-0.1** | **4.38e+10** | **127** |
| **Voters** | **1.69e-07** | **7.43e+03** | **127** |
| TotalVotes | 2.36e-12 | 1.03e-01 | 127 |
| LargestShare | 8.85e-13 | 3.88e-02 | 127 |
| Order | 5.00e-13 | 2.19e-02 | 127 |
| Panel B: Measurements related to MKR balance | | | |
| | Explained Variance Ratio | Explained Variance | N of obs. |
| **10k-100k** | **8.98e-01** | **1.58e-02** | **810** |
| **>100k** | **1.02e-01** | **1.79e-03** | **810** |
| >10k | 4.61e-18 | 8.10e-20 | 810 |

Note: This table reports the results for Principal Component Analysis (PCA) of centralization measurements. For each measurement of governance centralization, we present the ratio of explained variance and the variance explained by this measurement, respectively. A higher 'explained variance ratio' implies that the measurement can capture more information included by all measurements. To simplify our empirical results, we will only present results related to the measurements with highest 'explained variance ratios'.

### 4.2 Second stage: Factor analysis

In this section, we will first apply a series of univariate regressions, with MKR and DAI used as dependent variables and the measurements of centralized governance as independent variables. We consider financial, transaction, network and Twitter sentiment factors. In other words, we estimate the following regressions:

$$factor_{i,t} = \beta_0 + \beta_1 central_t + \varepsilon_t \quad (11)$$

Where:

- $i = \{MKR, DAI\}$
- $central_t = \{Voters_t, Gini_t, 10k - 100k_t, > 100k_t, Delegate_t\}$

Given $i$, factors can be defined as a set:

$$factor_{i,t} = \{financial_{i,j,t}, transaction_{i,k,t}, network_{i,l,t}, Twitter\ sentiment_{i,n,t}\}$$

Where $j = 1, ..., 7, k = 1, ..., 7, l = 1, ..., 4, m = 1, ..., 4$, and $n = 1, ..., 3$.

The detailed description of the above set of factors[8] is presented in detail in the Appendix A.1. For all statistically significant results, we run Granger test to address the problems of reverse causality. Appendix A.2 reports the results, and we discard the suspicious empirical findings.

### 4.2.1 Financial factors

Maker governance polls are directly related to non-technical changes, e.g., adding a new collateral, to Maker protocol. These changes will add more financial functions or revise the parameters of transactions on the protocol. Therefore, it is crucial to examine whether MKR and DAI factors, such as daily return, market cap, and trading volume are going to be affected by centralized governance, in the form of the metrics discussed in section 3. The univariate regression findings for these factors are summarized in table 7.

The two panels of the table bring forward some interesting findings for the effects of centralized governance measures for MKR and DAI. We observe that $10k - 100k$ and $delegate$ have a significant positive effect in the MKR volume, while $> 100k$ has the opposite. This conceptually means that, centralized voting power of major MKR holders and delegates will booster trading activities of MKR. On the other hand, if whales accumulate more

---
[8] To avoid the problems of spurious regressions, we first examine if factors are stationary. For the non-stationary variables, we choose the first differences of the variables instead. More details are given in online appendix 2.

MKR balances, both total volume and volume of large transactions will decrease. This could be a finding towards the claimed value proposition of MakerDAO, i.e., decentralized governance. The above could create a parallel with the findings of Meirowitz and Pi (2022) who analyze the shareholder's dilemma through the lens of voting and trading. Our results further imply that voting and trading are not substitute in some cases, and large stakeholders can have dissimilar influences. In the context of DAO, major holders and whales can even have contrary effects on trading volume of governance tokens.

Table 7. Financial factors (MKR, DAI)

| | | PANEL A: MKR | | | |
|---|---|---|---|---|---|
| | Voters | Gini | 10k-100k | >100k | Delegate |
| Return | -0.02 | 0.02 | 0.01 | -0.01 | -0.02 |
| | (-0.64) | (1.43) | (1.19) | (-1.06) | (-0.72) |
| $\Delta$MktC | -0.01 | 0.00 | 0.00 | 0.00 | 0.00 |
| | (-1.26) | (-0.19) | (0.53) | (-0.44) | (-0.21) |
| **Volume** | 0.00 | 0.00 | **0.09*** | **-0.07*** | -0.04 |
| | (0.45) | (0.22) | **(6.56)** | **(-6.40)** | (-1.14) |
| **Volume_usd** | 0.01 | 0.00 | **0.11*** | **-0.12*** | **0.13*** |
| | (0.79) | (0.47) | **(9.19)** | **(-13.03)** | **(3.20)** |
| **Volume_l** | 0.12 | -0.04 | **0.11*** | **-0.09*** | -0.03 |
| | (0.82) | (-0.76) | **(6.68)** | **(-6.88)** | (-0.74) |
| **Volume_l_usd** | 0.10 | -0.02 | **0.09*** | **-0.10*** | **0.11*** |
| | (1.13) | (-0.70) | **(8.32)** | **(-11.95)** | **(3.00)** |
| | | PANEL B: DAI | | | |
| | Voters | Gini | 10k-100k | >100k | Delegate |
| **Price** | **-0.02*** | 0.01 | **-0.11*** | **0.09*** | 0.00 |
| | **(-2.20)** | (1.13) | **(-9.41)** | **(10.23)** | (-0.28) |
| $\Delta$Return | 0.00 | 0.00 | 0.00 | 0.00 | 0.00 |
| | (0.33) | (0.17) | (-0.63) | (0.55) | (0.36) |
| **$\Delta$MktC** | 0.00 | 0.00 | **0.03*** | **-0.04*** | 0.00 |
| | (0.00) | (-0.08) | **(3.40)** | **(-4.94)** | (-0.05) |
| **Volume** | 0.02* | 0.01 | **0.05*** | **-0.05*** | 0.03 |
| | (1.68) | (0.82) | **(6.70)** | **(-9.55)** | (1.29) |
| **Volume_usd** | 0.02* | 0.01 | **0.05*** | **-0.05*** | 0.03 |
| | (1.68) | (0.83) | **(6.68)** | **(-9.75)** | (1.49) |
| **Volume_l** | 0.02* | 0.01 | **0.04*** | **-0.05*** | 0.03 |
| | (1.71) | (0.79) | **(6.28)** | **(-9.07)** | (1.29) |
| **Volume_l_usd** | 0.02* | 0.01 | **0.04*** | **-0.05*** | 0.03 |
| | (1.70) | (0.80) | **(6.24)** | **(-9.01)** | (1.29) |

Note: This table reports the univariate regression coefficients and standard t- statistics in the parentheses for the case of the financial factors of MKR (Panel A) and DAI (Panel B). *, **, and *** denote significance levels at the 10%, 5%, and 1% levels, respectively. The definitions of the factors are given in Table A.1. Only results without problems of reverse causality are in bold.

For the case of DAI, we continue to observe significant effects of centralized governance. $Voters$ leads to a significant decrease of the DAI price. Given that DAI is a stablecoin, decreasing price can be a signal of depegging, implying that decentralized governance can cause problems related to DAI. Though the fundamental goal of stablecoins is price stability, decreasing price can be the tip of the iceberg. For example, when the Luna crypto network collapsed[9], we witnessed that the two tokens, namely Luna and TerraUSD, went to zero, and the collapse caused unacceptable losses (approximately $60 billion) of the investors.

DAI's market cap is expected to be a metric of performance of Maker protocol. We observe that $10k - 100k$ is positively related to $\Delta MktC$, while $> 100k$ has the opposite effect. Regarding trading volume of DAI, $10k - 100k$ and $> 100k$ show inverse influences as well. Therefore, though both major holders and whales have centralized voting power (compared with small MKR holders), their influences can be different.

All the above show that centralized governance is significantly evident for the financial factors relevant to MKR and DAI. Especially, the trading volumes of MKR and DAI are driven by the proportion of MKR held by large holders. Generally, both sets of results bring forward a trade-off between decentralized governance and volume of these two tokens, and we also contend that the influences of dominant decision-makers can be complicated. Although the financial characteristics of coins are quite important, the literature has shown that the researcher needs to expand on other indicators integral to the technical structure of coins, tokens and protocols in order to capture their potential fundamental value (Kraaijeveld & De Smedt, 2020; Nadler & Guo, 2020; Liu & Tsyvinski, 2021; Nakagawa & Sakemoto, 2022). For that reason, we next expand in other MKR and DAI related factors that are non-financial in nature to further establish that centralized governance exists in the Maker protocol.

---

[9] https://www.forbes.com/sites/qai/2022/09/20/what-really-happened-to-luna-crypto/?sh=37d207874ff1

### 4.2.2 Transaction factors

Centralized governance should manifest also in the activities of MKR and DAI traders. These activities are captured by transaction factors, such as the average transaction size, the number of transactions, transaction volume on crypto exchanges, and transaction volume on *Decentralized Exchanges (DEXes)*. As we did with the financial factors, we present the univariate regression findings for these transaction indicators in the following table.

Table 8. Transaction factors (MKR, DAI)

| | Voters | Gini | 10k-100k | >100k | Delegate |
|---|---|---|---|---|---|
| **PANEL A: MKR** | | | | | |
| **AvgSizeMkr** | 0.05 | -0.03 | 0.01 | -0.02** | -0.02 |
| | (0.44) | (-0.92) | (1.45) | (-2.02) | (-0.69) |
| **AvgSize_usd** | 0.11 | -0.02 | **0.04*** | **-0.05*** | 0.03 |
| | (1.40) | (-0.77) | **(6.06)** | **(-10.18)** | (1.32) |
| **TxnCnt** | 0.10*** | -0.01 | **0.16*** | **-0.11*** | -0.06 |
| | (2.58) | (-0.60) | **(11.14)** | **(-9.65)** | (-1.28) |
| **Volume_ex** | 0.05 | 0.00 | **0.10*** | **-0.08*** | 0.03 |
| | (1.39) | (-0.18) | **(7.71)** | **(-8.95)** | (0.82) |
| **Volume_ex_usd** | 0.03** | 0.00 | **0.08*** | **-0.08*** | **0.12*** |
| | (2.18) | (-0.17) | **(8.91)** | **(-12.76)** | **(4.16)** |
| **Volume_dex** | 0.02 | 0.01 | **0.12*** | **-0.10*** | 0.01 |
| | (0.45) | (0.29) | **(7.83)** | **(-8.60)** | (0.27) |
| **Volume_dex_usd** | 0.03 | 0.01 | **0.12*** | **-0.12*** | **0.18*** |
| | (0.68) | (0.28) | **(9.38)** | **(-13.52)** | **(3.99)** |
| **PANEL B: DAI** | | | | | |
| **AvgSizeDai** | 0.02 | 0.00 | 0.01 | -0.02** | -0.02 |
| | (1.60) | (0.59) | (1.45) | (-2.02) | (-0.69) |
| **AvgSize_usd** | 0.02 | 0.00 | **0.04*** | **-0.05*** | 0.03 |
| | (1.58) | (0.62) | **(6.06)** | **(-10.18)** | (1.32) |
| **TxnCnt** | 0.09 | 0.01 | **0.16*** | **-0.11*** | -0.06 |
| | (1.47) | (0.34) | **(11.14)** | **(-9.65)** | (-1.28) |
| **Volume_ex** | 0.02 | 0.04 | **0.10*** | **-0.08*** | 0.03 |
| | (0.23) | (0.92) | **(7.71)** | **(-8.95)** | (0.82) |
| **Volume_ex_usd** | 0.01 | 0.04 | **0.08*** | **-0.08*** | **0.12*** |
| | (0.22) | (0.92) | **(8.91)** | **(-12.76)** | **(4.16)** |
| **ΔVolume_dex** | 0.02 | 0.01 | **0.12*** | **-0.10*** | 0.01 |
| | (0.58) | (0.64) | **(7.83)** | **(-8.60)** | (0.27) |
| **ΔVolume_dex_usd** | 0.02 | 0.01 | **0.12*** | **-0.12*** | **0.18*** |
| | (0.58) | (0.63) | **(9.38)** | **(-13.52)** | **(3.99)** |

Note: This table reports the univariate regression coefficients and standard t- statistics in the parentheses for the case of the transaction factors of MKR (Panel A) and DAI (Panel B). *, **, and *** denote significance levels at the 10%, 5%, and 1% levels, respectively. The definitions of the factors are given in Table A.2. Only results without problems of reverse causality are in bold.

The regression findings are again showing the significant effects of centralized governance in the trader activities across MKR and DAI investors. For example, we observe that $10k - 100k$ has a significant positive effect on the average transaction size in USD, volume on crypto exchanges (for both MKR and USD) and volume on DEXes (for both MKR and USD). However, $> 100k$ is negatively related to transaction factors of MKR, implying that the existence of whales can restrain MKR transactions. Additionally, $Voters$ can also lead to a higher volume (in USD) of MKR transactions on crypto exchanges, and $delegate$ can increase volumes (in USD) on both crypto exchanges and DEXes. In the case of DAI, the results are similar for the case of $10k - 100k$, $> 100k$ and $delegate$. The only difference is that $> 100k$ can decrease the average transaction size in DAI.

The main message of these findings is that there is substantial evidence of centralized governance effects in multi-faceted aspects of trading activities of MKR and DAI traders. Conceptually, it appears that large MKR holders can affect transaction volume and transaction size. The above could create a parallel to the traditional corporate finance inefficiencies suggested by studies such as Brav and Mathews (2011), Li et al. (2022) and Meirowitz and Pi (2022), namely that high trading volume can be observed after voting or shareholder meetings and that shareholders losing in the voting process, they may reduce their holdings. From this aspect, our results provide new insights: the influences of large stakeholders on trading volume do not only exist in a short period after voting or meeting.

Some early studies of corporate finance (e.g., Jensen and Warner, 1988; Leech, 1988) insist that voting power distribution can affect corporate performance, and recent work (e.g., Hein and Van Treeck, 2010) attempt to investigate how rising shareholder power relate to firm-specific factors and macroeconomic effects. Our research follows their intuition and proves that decision-makers in DAO (i.e., governance token holders) are crucial for the performance of DAO and the underlying DeFi protocol. Moreover, in the context of DAO, large stakeholders may affect a DAO very differently, which may inspire studies of corporate finance.

**4.2.3 Network and Twitter sentiment factors**

In this subsection, we focus on factors capturing network characteristics and social media sentiment. Firstly, as decentralized governance is the main selling point of Maker protocol, network statistics, such as total addresses, new addresses, active addresses, and their active ratio, may be differentially affected by centralized governance measurements. This is examined by the univariate regressions presented in table 9.

It is clear that centralized voting is affecting the MKR network factors. For MKR, $Voters$ exhibits significant positive effects on the total addresses with MKR balance, new addresses and active addresses, while $> 100k$ and $Delegate$ have the opposite effects. $Gini$ can also decrease the number of total addresses. Therefore, centralized voting seems to be negative. However, $10k - 100k$ can increase MKR network factors, so a degree of centralized voting power of major holders (but not whales) may be advantageous. To summarize, network adoption of DeFi protocol generally benefits from decentralized DAO governance, though centralization of MKR holding is not an exclusively deleterious thing.

For DAI, we observe the influence of $Voters$ and $Gini$. More $Voters$ will restrain the growth of new addresses, while $Gini$ can increase the growth of active ratio. The findings here again prove our arguments about the complex influences of centralized governance on network factors.

Table 9. Network factors (MKR, DAI)

| | Voters | Gini | 10k-100k | >100k | Delegate |
|---|---|---|---|---|---|
| **PANEL A: MKR** | | | | | |
| **TotalWithBlc** | 0.57*** | -0.18** | 0.12*** | -0.07*** | -0.13*** |
| | (3.15) | (-1.96) | (10.89) | (-8.64) | (-4.94) |
| **New** | 0.22*** | -0.04 | 0.24*** | -0.14*** | -0.31*** |
| | (4.02) | (-1.32) | (13.68) | (-10.03) | (-6.73) |
| **Active** | 0.21*** | -0.04 | 0.23*** | -0.16*** | -0.21*** |
| | (4.03) | (-1.46) | (14.39) | (-12.33) | (-4.52) |
| **ΔActiveRatio** | -0.02 | 0.03* | 0.00 | 0.00 | 0.00 |
| | (-0.52) | (1.75) | (0.02) | (0.05) | (0.19) |
| **PANEL B: DAI** | | | | | |
| | Voters | Gini | 10k-100k | >100k | Delegate |
| ΔTotalWithBlc | -0.02 | 0.00 | 0.00 | 0.00 | 0.00 |
| | (-0.53) | (-0.20) | (0.14) | (0.10) | (-0.08) |
| ΔNew | **-0.08**** | 0.00 | 0.00 | 0.00 | -0.01 |
| | **(-1.98)** | (-0.04) | (-0.14) | (0.28) | (-0.16) |
| ΔActive | -0.05 | -0.03 | 0.00 | 0.00 | -0.01 |
| | (-1.11) | (-1.02) | (-0.09) | (0.34) | (-0.17) |
| ΔActiveRatio | -0.03 | **0.04**** | 0.00 | 0.00 | 0.00 |
| | (-0.84) | **(2.22)** | (-0.14) | (0.14) | (0.01) |

Note: This table reports the univariate regression coefficients and standard t- statistics in the parentheses for the case of the network factors of MKR (Panel A) and DAI (Panel B). *, **, and *** denote significance levels at the 10%, 5%, and 1% levels, respectively. The definitions of the factors are given in Table A.3. Only results without problems of reverse causality are in bold.

In terms of social media sentiment, we focus on Twitter. Currently, Twitter is the main social media platform, where DeFi investors express their opinions. The community can discuss and communicate with each other, while the Maker protocol also has official accounts on Twitter. Focusing on Twitter sentiment factors, we are able to investigate if centralization in governance affects the sentiment tone of users. These results are presented in the following table.

Table 10. Twitter sentiment factors (MKR, DAI)

| PANEL A: MKR | | | | | |
|---|---|---|---|---|---|
| | Voters | Gini | 10k-100k | >100k | Delegate |
| Positive | 0.05 | -0.01 | **0.13**** | **-0.16**** | **0.17**** |
| | (0.82) | (-0.23) | **(9.00)** | **(-15.01)** | **(4.71)** |
| Neutral | 0.01 | 0.00 | **0.08**** | **-0.11**** | **0.11**** |
| | (0.26) | (-0.02) | **(5.78)** | **(-11.10)** | **(3.67)** |
| Negative | -0.01 | -0.02 | **0.05**** | **-0.06**** | **0.16**** |
| | (-0.12) | (-0.69) | **(3.06)** | **(-5.05)** | **(5.39)** |
| PANEL B: DAI | | | | | |
| | Voters | Gini | 10k-100k | >100k | Delegate |
| Positive | **0.13**** | -0.02 | **0.06**** | **-0.08**** | **0.16**** |
| | **(2.72)** | (-0.72) | **(4.96)** | **(-8.63)** | **(5.29)** |
| Neutral | -0.01 | 0.00 | 0.00 | **-0.01**** | -0.01 |
| | (-0.81) | (0.19) | (0.23) | **(-2.54)** | (-0.30) |
| Negative | 0.00 | 0.02 | 0.00 | **-0.04**** | **0.21**** |
| | (0.05) | (0.60) | (0.15) | **(-3.92)** | **(9.58)** |

Note: This table reports the univariate regression coefficients and standard t- statistics in the parentheses for the case of the Twitter sentiment factors of MKR (Panel A) and DAI (Panel B). *, **, and *** denote significance levels at the 10%, 5%, and 1% levels, respectively. The definitions of the factors are given in Table A.4. Only results without problems of reverse causality are in bold.

As in previous factors, significant effects are identified for both cases. For MKR, $10k - 100k$ and $Delegate$ have a positive effect towards the three categories of tweets, suggesting that voting concentration encourages more discussion. However, $> 100k$ is negatively related to twitter discussion, implying that the existence of MKR whales may make a decentralized platform less attention-getting.

For DAI, $> 100k$ shows its negative influence on Twitter discussion again. $Voters$, $10k - 100k$, and $Delegate$ can increase the number of positive tweets, though $Delegate$ can also bring more negative tweets. $Gini$ surprisingly does not appear to provide any kind of sentiment attention. Not very surprisingly, more voters, as a signal of decentralized governance, can lead to more positive tweets. But the complex effects of delegates' voting power show that centralized governance results in more debate. Higher voting power of delegates can lead to more negative tweets than positive ones. The above effects of centralized measures in social media sentiment are another signal that decentralization in MakerDAO is not ensured and can bring debates on Twitter.

### 4.2.4 Collateral ratios

Finally, we investigate how centralized governance affects collateral assets locked in Maker protocol. To initiate loans from Maker protocol, Maker users must lock collaterals. Therefore, collateral assets accepted by Maker protocol are a weighty issue in Maker governance. Moreover, if the locked collateral assets become risky, Maker protocol may be affected. In March 2023, DAI was severely influenced by the depegging of USD Coin (USDC)[10], since USDC is one of the most important collateral assets in Maker protocol.

In this subsection, we examine how centralized governance relates to components of collateral assets. We first consider three categories of collateral assets, including Ether (ETH), Wrapped Bitcoin (WBTC), and stablecoins. The reason for including ETH and WBTC is intuitive. ETH, as the native cryptocurrency of Ethereum blockchain, was one of the earliest accepted collateral assets in Maker protocol. WBTC can be regarded as Bitcoin traded on Ethereum, and it also accounts for a significant proportion. Besides, stablecoins play a crucial role in Maker protocol. For each category of collateral assets, we compute the proportion of its value (in USD) to the total value of locked collateral assets in Maker protocol. Figure 6 shows that ETH was the dominant collateral before September 2020. Since September 2020, ETH ratio has been much lower, while stablecoins become important collaterals. After September 2020, stablecoins usually accounted for more than 25% of value of total collateral assets, and the spikes of stablecoin ratio are more than 50%.

**Figure 6. Proportions of three different collateral assets**

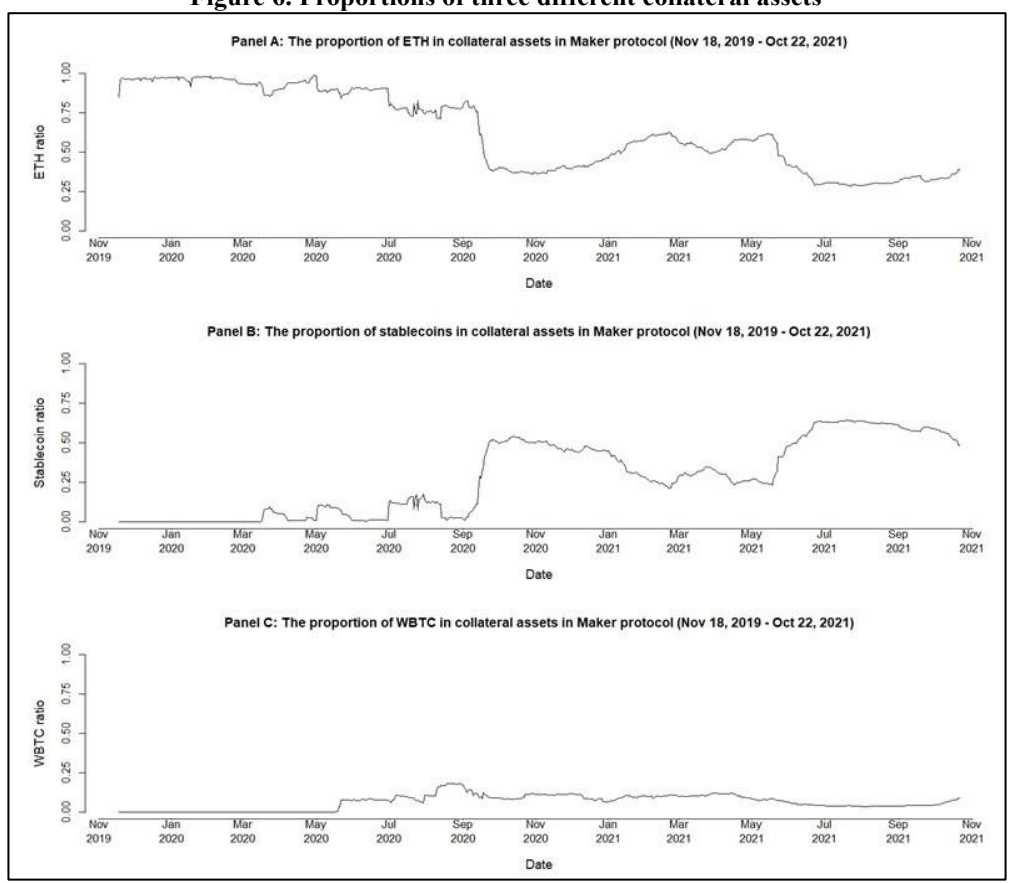

Note: This figure illustrates the proportions of three different collateral assets locked in Maker protocol (Nov 18, 2019 – Oct 22, 2021), including ETH, stablecoins and WBTC. The datasets are queried from *dune.xyz*.

To explore how collateral assets are driven by centralized governance, we estimate the following regressions:

---

[10] https://cointelegraph.com/news/maker-dao-files-emergency-proposal-addressing-3-1b-usdc-exposure

$$Collateral_t = \beta_0 + \beta_1 central_t + \varepsilon_t \quad (12)$$

Where:

- $central_t = \{Voters_t, Gini_t, 10k-100k_t, >100k_t, Delegate_t\}$

Given $i$, factors can be defined as a set:

$$Collateral_t = \{\Delta ETH\ ratio_t,\ \Delta Stablecoin\ ratio_t, \Delta WBTC\ ratio_t\}$$

The detailed description of the above set of factors is presented in detail in the table A.7 in Appendix A.1. For all statistically significant results, we run Granger test to address the problems of reverse causality. Table A.13 in Appendix A.2 reports the results, and we discard the suspicious empirical findings.

The picture here is clear. Distribution of MKR is related to collateral ratios in Maker protocol. Increased voting power of $>100k$ and $Delegate$ can decrease the growth of ETH ratio, while $10k-100k$ has the opposite effect. For stablecoins, $10k-100k$ and $Delegate$ show the contrary influences as well. Again, centralized voting power can affect Maker protocol from the aspect of collateral assets, given that Maker governance decides collateral onboarding and offboarding. Furthermore, MKR holders have dissimilar preferences of collaterals, which may explain why major holders and whales show different influences. Overall, our two-stage analysis is providing substantial empirical evidence towards centralized governance in MakerDAO, as several significant univariate relationships are established across different classes of factors[11]. The next section is providing some further robustness checks towards that end.

Table 11. Collateral ratios

|  | Voters | Gini | 10k-100k | >100k | Delegate |
|---|---|---|---|---|---|
| ΔETH_ratio | 0.02 | -0.02 | **0.02**** | **-0.01*** | **-0.03**** |
|  | (0.60) | (-1.15) | **(2.04)** | **(-1.73)** | **(-2.88)** |
| ΔStablecoin_ratio | -0.01 | 0.02 | **-0.02**** | 0.01* | **0.04**** |
|  | (-0.29) | (0.85) | **(-1.96)** | (1.82) | **(3.03)** |
| ΔWBTC_ratio | -0.02 | 0.03 | **-0.02**** | **-0.02**** | -0.02 |
|  | (-0.53) | (1.23) | **(-1.99)** | **(-1.99)** | (-1.31) |

Note: This table reports the univariate regression coefficients and standard t- statistics in the parentheses for the case of the collateral ratios in Maker protocol. *, **, and *** denote significance levels at the 10%, 5%, and 1% levels, respectively. The definitions of the factors are given in Table A.5 in Appendix 1. Only results without problems of reverse causality are in bold.

## 5. Robustness checks

### 5.1 Addressing endogeneity: Off-chain governance as an instrumental variable (MKR)

The empirical results presented in section 4 could face criticism due to potential endogeneity concerns. To alleviate this issue, we use the instrumental variable approach and estimate two-stage least squares (2SLS) regressions. We construct an instrumental variable (IV) using datasets for forum signal threads, that are a part of the off-chain governance in the Maker protocol. Anyone can participate in the discussion and voting in the threads. That means that, unlike the on-chain governance investigated previously, off-chain governance does not require MKR in one's account. Even people who do not use blockchain can share their opinions and click an option in the thread. For some signal threads, the informal discussion will finally turn to Maker IP, where voters will choose their options.

---

[11] Given the extent of factors and univariate regressions examined, we also present summary of the relationships in online appendix OA.3.

The results of signal threads could be thought are related with on-chain governance. On the other hand, as a non-custodial stablecoin, the price of DAI is independent of Maker protocol but depends on exogenous collateral (Klages-Mundt et al., 2020). So, the direct relationship between DAI and off-chain governance is not well understood. Similarly, off-chain governance may not be an endogenous variable to other factors of Maker protocol. Theoretically, Maker governors aim for maximizing their revenue (Klages-Mundt & Minca, 2022). In other words, the factors, e.g., transaction factors, are outcomes of decisions jointly made by voters in Maker governance. Finally, our IV does not suffer from reverse causality when estimating the first-stage regressions. Undoubtedly, Maker users could have more discussion when the protocol performs well or badly, but their debates are usually observed in other categories of threads. Functionally, forum signal threads are warm-up of on-chain voting, so off-chain voting in these threads affects Maker protocol via the following on-chain voting.

For each thread, we document the first post date and the number of voters. In some threads, there will be several voting polls, and we will only count unique voters within a thread. Then, the number of off-chain voters daily can be calculated. This is a valid instrument for our setup as explained in the online appendix OA.4. Based on the above we report the first and the second stage regressions for the case of the financial and transaction factors of MKR in the following two tables[12].

Table 12: 2-SLS IV regressions (financial factors – MKR)

| Panel A: Estimate *Voters* using an instrument | | | | | | | | | |
|---|---|---|---|---|---|---|---|---|---|
| | (1) | (2) Return | (3) Volume | (4) Volume_usd | (5) | (6) ΔMktC | (7) | (8) Volume_l | (9) Volume_l_usd |
| Off-chain | 0.17** (4.90) | | | | 0.18** (5.02) | | 0.17* (2.86) | | |
| Voters | | -0.21 (-1.62) | 0.02 (0.94) | 0.14 (1.43) | | -0.47 (-1.14) | | 0.90 (1.05) | 0.25 (0.68) |
| Durbin's test | | 1.90 | 0.51 | 1.58 | | 1.90 | | 1.10 | 0.11 |
| p-value | | 0.17 | 0.48 | 0.21 | | 0.17 | | 0.29 | 0.74 |
| Wu-Hausman test | | 1.89 | 0.50 | 1.56 | | 1.88 | | 1.07 | 0.10 |
| p-value | | 0.17 | 0.48 | 0.21 | | 0.17 | | 0.31 | 0.75 |
| Adj. R-sq | | -0.51 | -0.14 | -0.42 | | -0.48 | | -0.43 | -0.04 |
| N | | 127 | 127 | 127 | | 126 | | 67 | 67 |
| Panel B: Estimate *10k-100k* using an instrument | | | | | | | | | |
| | (1) | (2) Return | (3) Volume | (4) Volume_usd | (5) ΔMktC | | (6) | (7) Volume_l | (8) Volume_l_usd |
| Off-chain | 0.32*** (6.02) | | | | | | 0.31** (5.81) | | |
| 10k-100k | | -0.19 (-0.68) | 0.22 (1.01) | 0.28** (2.03) | -0.26 (-1.30) | | | 0.16 (0.63) | 0.24* (1.64) |
| Durbin's test | | 0.26 | 0.20 | 0.08 | 2.68 | | | 0.00 | 0.00 |
| p-value | | 0.61 | 0.65 | 0.78 | 0.10 | | | 0.95 | 0.95 |
| Wu-Hausman test | | 0.26 | 0.20 | 0.08 | 2.67 | | | 0.00 | 0.00 |
| p-value | | 0.61 | 0.66 | 0.78 | 0.10 | | | 0.95 | 0.95 |
| Adj. R-sq | | -0.03 | -0.01 | 0.16 | -0.36 | | | 0.02 | 0.16 |
| N | | 126 | 126 | 126 | 126 | | | 127 | 127 |

Note: This table reports results of the 2-SLS IV regressions. Panel A, Columns (1), (5) and (7) report the results of the following first stage regression: $Voters_t = \beta_0 + \beta_1 off-chain_t + \varepsilon_t$, where $off-chain$ is an instrumental variable. Columns (2) – (4), (6) and columns (8) – (9) report the results of second stage: $factor_t = \beta_0 + \beta_1 \widehat{Voters}_t + \varepsilon_t$. In Columns (1), (5) and (7), partial F-statistics are reported in parentheses. In Columns (2) - (4), (6), and (8) - (9), t-statistics are reported in parentheses. *, **, and *** denote significance levels at the 10%, 5%, and 1%, respectively. Panel B, Columns (1) and (6) report the results of the following first stage regression: $10k-100k_t = \beta_0 + \beta_1 off-chain_t + \varepsilon_t$, where $off-chain$ is an instrumental variable. Columns (2) – (5) and columns (7) – (8) report the results of second stage: $factor_t = \beta_0 + \beta_1 \widehat{10k-100k}_t + \varepsilon_t$. In Columns (1) and (6), partial F-statistics are reported in parentheses. In Columns (2) - (5) and (7)-(8), t-statistics are reported in parentheses. *, **, and *** denote significance levels at the 10%, 5%, and 1%, respectively.

Table 13: 2-SLS IV regressions (transaction factors – MKR)
Panel A: Estimate *Voters* using an instrument

---

[12] For the sake of space, the remaining 2SLS results for MKR and all the equivalent 2SLS results for DAI are provided also in the online appendix OA.4.

|  | (1) | (2) AvgSize Mkr | (3) AvgSize usd | (4) | (5) TxnC nt | (6) Volume ex | (7) Volume ex_usd | (8) | (9) Volume dex | (10) Volume dex_usd |
|---|---|---|---|---|---|---|---|---|---|---|
| Off-chain | 0.17* (2.86) |  |  | 0.17** (4.90) |  |  |  | 0.18** (5.02) |  |  |
| Voters |  | 0.16 (0.38) | 0.21 (0.59) |  | 0.61** (2.43) | 0.02 (0.94) | 0.14 (1.43) |  | 0.80** (1.94) | 0.66 (1.56) |
| Durbin's test |  | 0.05 | 0.06 |  | 4.95 | 0.51 | 1.58 |  | 6.59 | 3.89 |
| p-value |  | 0.82 | 0.80 |  | 0.03 | 0.48 | 0.21 |  | 0.01 | 0.05 |
| Wu-Hausman test |  | 0.05 | 0.06 |  | 5.03 | 0.50 | 1.56 |  | 6.78 | 3.92 |
| p-value |  | 0.82 | 0.81 |  | 0.03 | 0.48 | 0.21 |  | 0.01 | 0.05 |
| Adj. R-sq |  | -0.03 | -0.01 |  | -1.21 | -0.14 | -0.42 |  | -1.71 | -1.01 |
| N |  | 67 | 67 |  | 127 | 127 | 127 |  | 126 | 126 |
| **Panel B: Estimate *10k-100k* using an instrument** | | | | | | | | | | |
|  | (1) | (2) AvgSizeMkr | (3) AvgSize usd | (4) Volume_dex | (5) Volume dex_usd | (6) | (7) TxnC nt | (8) Volume ex | (9) Volume ex_usd |  |
| Off-chain | 0.31** (5.81) |  |  |  |  | 0.32*** (6.02) |  |  |  |  |
| 10k-100k |  | -0.11 (-0.50) | 0.15 (0.81) | 0.45*** (2.47) | 0.38** (2.12) |  | 0.77*** (2.92) | 0.19 (1.31) | 0.40*** (2.76) |  |
| Durbin's test |  | 0.05 | 0.33 | 4.31 | 1.53 |  | 3.04 | 0.25 | 0.44 |  |
| p-value |  | 0.83 | 0.57 | 0.03 | 0.22 |  | 0.08 | 0.61 | 0.51 |  |
| Wu-Hausman test |  | 0.05 | 0.32 | 4.36 | 1.51 |  | 3.04 | 0.25 | 0.43 |  |
| p-value |  | 0.83 | 0.57 | 0.04 | 0.22 |  | 0.08 | 0.62 | 0.51 |  |
| Adj. R-sq |  | -0.01 | 0.13 | -0.51 | -0.07 |  | -0.22 | 0.00 | 0.17 |  |
| N |  | 127 | 127 | 127 | 127 |  | 126 | 126 | 126 |  |

Note: This table reports results of the 2-SLS IV regressions. Panel A, Columns (1), (4) and (8) report the results of the following first stage regression: $Voters_t = \beta_0 + \beta_1 off-chain_t + \varepsilon_t$, where $off-chain$ is an instrumental variable. Columns (2) – (3), (5) – (7) and (9) - (10) report the results of second stage: $factor_t = \beta_0 + \beta_1 \widehat{Voters}_t + \varepsilon_t$. In Columns (1), (4) and (8), partial F-statistics are reported in parentheses. In Columns (2) - (3), (5)-(7), and (9)-(10), t-statistics are reported in parentheses. *, **, and *** denote significance levels at the 10%, 5%, and 1%, respectively.

Panel B, Columns (1) and (6) report the results of the following first stage regression: $10k - 100k_t = \beta_0 + \beta_1 off - chain_t + \varepsilon_t$, where $off - chain$ is an instrumental variable. Columns (2) – (5) and columns (7) – (9) report the results of second stage: $factor_t = \beta_0 + \beta_1 \widehat{10k - 100k}_t + \varepsilon_t$. In Columns (1) and (6), partial F-statistics are reported in parentheses. In Columns (2) - (5) and (7)-(9), t-statistics are reported in parentheses. *, **, and *** denote significance levels at the 10%, 5%, and 1%, respectively.

Before using an instrument, we first test if our instrument suffers from weak instruments concerns. The results for the first stage regressions show that our instrument can be used in 2SLS regressions. Then, we are curious that if measurements of centralized governance are endogenous to factors for MKR and DAI. To test the endogeneity, we apply Durbin's test and Wu-Hausman test. Since the null hypothesis is that endogeneity does not exist, usually, we do not observe endogeneity between our measurements and factors for MKR and DAI, meaning that the corresponding OLS regressions in section 4 are reliable. For results where endogeneity is observed (e.g., Column (8), Panel B of Table 12), we compare results of 2SLS regressions and results in section 4, and the findings are consistent. Therefore, the measurements of centralized governance are generally not found endogenous to MKR and DAI factors.

### 5.3 Regression continuity

Governance centralization may be less severe in a specific period of time. For example, figure 3 shows that the largest voters' voting share is not very high from poll #300 and poll #500. Therefore, regression continuity is a potential problem of our empirical results. To address this concern, we estimate the univariate regressions in section 4 with a subset of Maker governance polls. The subset includes governance polls from poll #298 (deployed on September 21, 2020) to poll #499 (deployed on March 15, 2021), and the table below summarizes the results based on this subset. Though there are some new statistically significant results (e.g., the relationship between *Voters* and daily return of MKR), most results for regression continuity analysis are consistent with our findings in section 4.

**Table 14: Regression continuity (summary)**

| Panel A: MKR | | | | |
|---|---|---|---|---|
| **Measurements** | Financial | Transaction | Network Factors | Twitter Sentiment Factors |

|  | factors | Factors |  |  |
|---|---|---|---|---|
| **Voters** | Return ↓ |  |  |  |
| **Gini** |  |  |  |  |
| **10k-100k** | Volume_usd↑<br>Volume_l_usd↑ | AvgSize_usd↑<br>TxnCnt↑<br>Volume_ex↑<br>Volume_ex_usd↑<br>Volume_dex↑<br>Volume_dex_usd↑ |  | Positive↑ |
| **>100k** | Volume_usd↓<br>Volume_l_usd↓ | AvgSize_usd↓<br>TxnCnt↓<br>Volume_ex↓<br>Volume_ex_usd↓<br>Volume_dex↓<br>Volume_dex_usd↓ |  | Positive↓ |
| **Delegate** |  |  |  |  |
| **Panel B: DAI** |  |  |  |  |
| **Measurements** | Financial factors | Transaction Factors | Network Factors | Twitter Sentiment Factors |
| **Voters** | ΔMktC ↓ |  |  | Positive↑ |
| **Gini** | ΔReturn ↑<br>Volume_usd ↑ |  | ΔTotalWithBlc ↑ |  |
| **10k-100k** | Price ↓<br>**ΔMktC**↑<br>**Volume**↑<br>**Volume_usd**↑<br>**Volume_l**↑<br>**Volume_l_usd**↑ | TxnCnt↑ |  |  |
| **>100k** | Price ↑<br>**ΔMktC**↓<br>**Volume**↓<br>**Volume_usd**↓<br>**Volume_l**↓<br>**Volume_l_usd**↓ | TxnCnt↓ |  |  |
| **Delegate** |  | ΔVolume_dex↑<br>ΔVolume_dex_usd↑ |  |  |

Note: This table reports the relationship between measurements of centralized voting power and the factors of MKR and DAI. The measurements are calculated using the datasets for governance polls deployed from September 21, 2020 to March 15, 2021. The detailed regression results are presented in online appendix OA.5. The results consistent with section 4 are in bold.

## 5.4 Certain type of governance polls

Governance polls can be categorized, and the labels are publicly observable on Maker governance forum. Generally, the labels reflect on the focus of a governance poll. For example, 'collateral onboarding' polls are about new collateral assets that can be used to initiate loans from Maker protocol, while 'MIP' polls discuss improvement proposals of Maker protocol. In the subsection, we focus on 'risk parameter', which has the most governance polls. Using the subset, we calculate *Voters* and *Gini* again and estimate the univariate regressions in section 4.

Interestingly, most findings in table 16 are different from our findings in section 4. Given that 'risk parameter' polls can decide key variables of Maker protocol, e.g., interest rates and debt ceilings, it is not very surprising that governance centralization in these polls can affect financial factors and transaction factors (for both MKR and DAI). More decentralized governance (e.g., more *Voters*) has positive effects (e.g., increased volume, more transactions), while higher *Gini* can restrain transactions, the transaction size, and trading volume.

Overall, this subsection shows that decentralized governance is advantageous. Further studies can combine the contents of governance polls in different categories with the performance of Maker protocol, and we will be able to learn which issues are more crucial. Besides, we can also investigate voters' voting patterns in different types of polls, and their private benefits may be revealed.

Table 15: Categories of Maker governance polls

| | Number of polls |
|---|---|
| **Risk Parameter** | 252 |
| **Ratification Poll** | 27 |
| **Inclusion Poll** | 70 |
| **Collateral Onboarding** | 50 |
| **Collateral Offboarding** | 2 |
| **Greenlight** | 146 |
| **Real World Asset** | 28 |
| **Misc Governance** | 18 |
| **Misc Funding** | 3 |
| **MakerDAO Open Market Committee** | 11 |
| **MIP** | 106 |
| **Budget** | 25 |
| **Oracle** | 38 |
| **System Surplus** | 6 |
| **DAI Direct Deposit Module** | 1 |
| **Multi-chain Bridge** | 1 |
| **Technical** | 17 |
| **Auction** | 20 |
| **Delegates** | 0 |
| **Peg Stability Module** | 11 |
| **Core Unit Onboarding** | 17 |
| **Dai Savings Rate** | 28 |
| **Black Thursday** | 4 |
| **Multi-Collateral DAI Launch** | 5 |
| **Prioritization Sentiment** | 2 |

Note: This table reports the number of governance polls (poll#16 – poll#638) in different categories. One poll can have multiple labels.

Table 16: Measurements of governance centralization based on 'risk parameter' polls (summary)

| MKR | | | | |
|---|---|---|---|---|
| **Measurements** | Financial factors | Transaction Factors | Network Factors | Twitter Sentiment Factors |
| **Voters** | Volume ↑ Volume_usd ↑ Volume_1_usd ↑ | AvgSizeMkr ↑ TxnCnt ↑ Volume_ex ↑ Volume_dex_usd ↑ | **TotalWithBlc**↑ **New**↑ **Active**↑ | Neutral ↓ |
| **Gini** | Volume_usd ↓ Volume_1_usd ↓ | AvgSize_usd ↓ TxnCnt ↓ Volume_ex_usd ↓ Volume_dex_usd ↓ | | |
| **Panel B: DAI** | | | | |
| **Measurements** | Financial factors | Transaction Factors | Network Factors | Twitter Sentiment Factors |
| **Voters** | | Volume_ex ↑ Volume_ex_usd ↑ ΔVolume_dex ↑ ΔVolume_dex_usd ↑ | ΔTotalWithBlc ↓ | Neutral ↓ |
| **Gini** | ΔReturn ↑ ΔMktC ↓ Volume_usd ↓ | | | |

Note: This table reports the relationship between measurements of centralized voting power and the factors of MKR and DAI. The measurements are calculated using the datasets for governance polls with the label 'risk parameter'. The detailed regression results are presented in online appendix OA.6.

## 6. Conclusion

Decentralization is crucial innovation of blockchain, and the rapid growth of DeFi relies on decentralization. Complete decentralization is theoretically impossible (Abadi & Brunnermeier, 2022) and empirical evidence of centralization is detected in different layers of blockchain (Sai et al., 2021). In this paper, we focus on governance in DeFi and particularly on the Maker protocol, which is governed by MakerDAO. Decentralized governance is a crucial domain for DeFi and Maker protocol is an ideal case since its voting history is considered transparent and precise (Beck et al., 2018). By examining Maker governance polls, we find that voters are centralized in a small group and voting power is unequally distributed among these voters. In most voting activities, the largest voters could account for a significant proportion of votes. Previously, Gervais et al. (2014) and Azouvi et al. (2018) argue that a few key developers have unilateral decision-making power in blockchain governance. This problem might derive from the requirement of programming skills. Our results expand the discussion to the token-weighted voting system in DeFi. Particularly in Maker, any MKR holder can easily participate in governance by clicking an option on the website, which would indicate that governance would be more decentralized. Interestingly, our results show that governance in Maker protocol is highly centralized.

To show that, we establish measurements of centralized governance in three categories, namely voting participation, centralized voting power, and distribution of governance token. By investigating the relationships between Maker-specific factors, we find that larger voting participation usually have positive influence on the Maker protocol. For example, transaction volume will be higher, and more users will be attracted. On the other hand, the high degree of centralized governance will refrain the protocol growth. In the case of DAI, the effects of larger voting participation may not be ideal. For example, more voters will cause lower price of DAI. Given that DAI is a stablecoin, decreasing price can be the start of depegging. Moreover, centralized voting power of large voters can contribute to more trading volume and transaction size, implying that centralization can be advantageous. Currently, the best governance structure for stablecoins is an open question. If governance is extremely centralized, several agents can easily manipulate the stablecoin systems (Hoang & Baur, 2022). This paper implies that centralized governance to some extent may contribute to trading activities of stablecoins. With our findings, we make a compelling case in favor of the argument that decentralization in DeFi platforms is an illusion and that the trade-off between market performance and decentralization is the rather true value proposition offered to investors in DeFi applications. This is a trade-off similar to the ones observed in the corporate world where unexpected results of governance processes may be caused by difference preferences of decision-makers (Garlappi et al., 2017; Donaldson et al., 2020).

Although our findings appear conceptually and empirically robust, they should be interpreted with their limitations in mind. First, the identity of dominant voters is unknown. Anonymity is another character of blockchain and DeFi, and we may not know voters' identity until they are willing to announce. Second, we do not illustrate the accumulation of significant voting power. After checking these dominant voters' transaction history, we find that most of them do not have many transactions of MKR. Few of them have transactions of other cryptocurrencies. In blockchain, any agent can have endless addresses, i.e., accounts. Therefore, these voters may have specific accounts for participating in Maker governance. Nadler & Schär (2020) propose mapping algorithms that can cluster several addresses owned by a single entity. Further studies can estimate governance centralization by applying similar methods. Finally, we do not know whether the authors of Maker IPs are dominant voters. If a dominant voter proposes changes to the Maker protocol, the aim of such proposals might be tied to the own vested interests. With their large voting power, this could lead to further centralization of power and potential collusion during the development of Maker protocol.

Currently, writing Maker IPs requires both programming skills and understanding of technical structure of DeFi. Assuming that not many voters have such competence, at least key developers can guide voters by proposing specific Maker IPs, implying that the centralized power of core developers exists in DeFi. This could also be supported by studies suggesting that delegating tasks to a group of experts can lead to better aggregation of information (Fehrler & Janas, 2021). As things stand, though, Maker users rely on developers to provide detailed proposals, the aims of codes and explanations of all possible outcomes in a more understandable way. Another way forward is the Maker protocol to make IP authors' addresses publicly available to allow users to detect suspicious activities of developers.

**References**

Abadi, J., & Brunnermeier, M. (2022). *Blockchain Economics* (NBER Working Paper No. 25407). National Bureau of Economic Research. https://www.nber.org/papers/w25407

Abdikerimova, S., & Feng, R. (2022). Peer-to-Peer multi-risk insurance and mutual aid. European Journal of Operational Research, 299(2), 735-749.

Anyfantaki, S., Arvanitis, S., & Topaloglou, N. (2021). Diversification benefits in the cryptocurrency market under mild explosivity. *European Journal Of Operational Research*, *295*(1), 378-393.

Aramonte, S., Huang, W., & Schrimpf, A. (2021). *DeFi risks and the decentralisation illusion* (BIS Qartely Review, December 2021). Bank for International Settlements. https://www.bis.org/publ/qtrpdf/r_qt2112b.htm.

Azouvi, S., Maller, M., & Meiklejohn, S. (2018). Egalitarian Society or Benevolent Dictatorship: The State of Cryptocurrency Governance. *Financial Cryptography And Data Security*, 127-143.

Bai, Q., Zhang, C., Xu, Y., Chen, X., & Wang, X. (2020). *Evolution of Ethereum: A Temporal Graph Perspective*. arXiv.org. Retrieved 29 August 2021, from https://arxiv.org/abs/2001.05251.

Bartoletti, M., Chiang, J., & Lluch-Lafuente, A. (2020). *SoK: Lending Pools in Decentralized Finance*. arXiv.org. Retrieved 29 August 2021, from https://arxiv.org/abs/2012.13230.

Börgers, C. (2010). Mathematics of social choice. Philadelphia, Pa.: Society for Industrial and Applied Mathematics.

Brams, S. (2008). Mathematics and Democracy. Princeton: Princeton University Press.

Brams, S., & Fishburn, P. (2002). Chapter 4 Voting procedures. *Handbook Of Social Choice And Welfare*, 173-236. Elsevier.

Beck, R., Müller-Bloch, C., & King, J. (2018). Governance in the Blockchain Economy: A Framework and Research Agenda. *Journal of the Association for Information Systems*, 19(10), 1020-1034.

Bollaert, H., Lopez-de-Silanes, F., & Schwienbacher, A. (2021). Fintech and access to finance. *Journal Of Corporate Finance*, *68*, 101941.

Brav, A., & Mathews, R. D. (2011). Empty voting and the efficiency of corporate governance. *Journal of Financial Economics*, *99*(2), 289–307. https://doi.org/10.1016/j.jfineco.2010.10.005

Carter, N., & Jeng, L. (2021, August 6). *DEFI protocol risks: The paradox of defi*. SSRN. Retrieved March 22, 2023, from https://papers.ssrn.com/sol3/papers.cfm?abstract_id=3866699

Choi, T. M., Guo, S., Liu, N., & Shi, X. (2020). Optimal pricing in on-demand-service-platform-operations with hired agents and risk-sensitive customers in the blockchain era. European Journal of Operational Research, 284(3), 1031-1042.

Cong, L., Li, Y., & Wang, N. (2020). Tokenomics: Dynamic Adoption and Valuation. *The Review Of Financial Studies*, *34*(3), 1105-1155.

Donaldson, J. R., Malenko, N., & Piacentino, G. (2020). Deadlock on the Board. *The Review of Financial Studies*, *33*(10), 4445–4488. https://doi.org/10.1093/rfs/hhaa006

Dorfman, R. (1979). A Formula for the Gini Coefficient. *The Review of Economics and Statistics*, 61(1), 146.

Fama, E., & Jensen, M. (1983). Separation of Ownership and Control. *The Journal Of Law And Economics*, *26*(2), 301-325.

Fehrler, S., & Janas, M. (2021). Delegation to a Group. *Management Science*, *67*(6), 3714-3743.

Garlappi, L., Giammarino, R. and Lazrak, A., 2017. Ambiguity and the corporation: Group disagreement and underinvestment. *Journal of Financial Economics*, 125(3), 417-433.

Gervais, A., Karame, G., Capkun, V., & Capkun, S. (2014). Is Bitcoin a Decentralized Currency?. *IEEE Security and Privacy*, *12*(3), 54-60.

Harvey, C.R., Ramachandran, A., & Santoro, J. (2021). *DeFi and the Future of Finance*. John Wiley & Sons.

Hein, E., & Van Treeck, T. (2010). Financialisation and rising shareholder power in Kaleckian/Post-kaleckian models of distribution and growth. *Review of Political Economy*, *22*(2), 205–233. https://doi.org/10.1080/09538251003665628

Hoang, L., & Baur, D. (2021). How stable are stablecoins?. *The European Journal Of Finance*. Advance online publication. https://doi.org/10.1080/1351847x.2021.1949369

Howell, S., Niessner, M., & Yermack, D. (2020). Initial Coin Offerings: Financing Growth with Cryptocurrency Token Sales. *The

*Review Of Financial Studies*, *33*(9), 3925-3974. https://doi.org/10.1093/rfs/hhz131

Hsieh, Y., (JP) Vergne, J., & Wang, S. (2017). The internal and external governance of blockchain-based organizations. *Bitcoin And Beyond*, 48-68. Taylor and Francis Group.

Jentzsch, C. (2016). Decentralized autonomous organization to automate governance [White paper]. https://lawofthelevel.lexblogplatformthree.com/wp-content/uploads/sites/187/2017/07/WhitePaper-1.pdf

Jensen, M. C., & Warner, J. B. (1988). The distribution of power among corporate managers, shareholders, and directors. *Journal of Financial Economics*, *20*, 3–24. https://doi.org/10.1016/0304-405x(88)90038-4

Jiang, S., Li, Y., Wang, S. and Zhao, L., 2022. Blockchain competition: The tradeoff between platform stability and efficiency. European Journal of Operational Research, 296(3), pp.1084-1097.

Karim, S., Lucey, B., Naeem, M., & Uddin, G. (2022). Examining the interrelatedness of NFTs, DeFi tokens and cryptocurrencies. *Finance Research Letters*, 102696.

Klages-Mundt, A., Harz, D., Gudgeon, L., Liu, J., & Minca, A. (2020). Stablecoins 2.0. *Proceedings Of The 2Nd ACM Conference On Advances In Financial Technologies*.

Kraaijeveld, O., & De Smedt, J. (2020). The predictive power of public Twitter sentiment for forecasting cryptocurrency prices. *Journal Of International Financial Markets, Institutions And Money*, *65*, 101188.

Klages-Mundt, A., & Minca, A. (2021). (In)Stability for the Blockchain: Deleveraging Spirals and Stablecoin Attacks. arXiv.org. Retrieved 11 February 2022, from https://arxiv.org/abs/1906.02152.

Lee, J. Y. (2019). A decentralized token economy: How blockchain and cryptocurrency can revolutionize business. Business Horizons, 62(6), 773-784.

Leech, D. (1988). The relationship between shareholding concentration and shareholder voting power in British companies: A study of the application of power indices for simple games. *Management Science, 34*(4), 509–527. https://doi.org/10.1287/mnsc.34.4.509

Li, S.Z., Maug, E. and Schwartz-Ziv, M., 2022. When shareholders disagree: Trading after shareholder meetings. *The Review of Financial Studies*, 35(4), pp.1813-1867.

Liu, Y., & Tsyvinski, A. (2020). Risks and Returns of Cryptocurrency. *The Review Of Financial Studies*, *34*(6), 2689-2727.

MakerDAO. (2020). The Maker Protocol White Paper [White paper]. https://makerdao.com/en/whitepaper

MakerDAO. (2021). A Guide to Participating in MakerDAO Governance. https://makerdao.world/en/learn/governance/how-voting-works/

Meirowitz, A. and Pi, S., 2022. Voting and trading: The shareholder's dilemma. *Journal of Financial Economics*, forthcoming.

Momtaz, P.P. (2022). How efficient is Decentralized Finance (DeFi)?. Retrieved 15 April 2022, from http://dx.doi.org/10.2139/ssrn.4063670.

Nadler, P., & Guo, Y. (2020). The fair value of a token: How do markets price cryptocurrencies?. *Research In International Business And Finance*, *52*, 101108.

Nadler, M., & Schär, F. (2020, December 16). *Decentralized Finance, centralized ownership? an iterative mapping process to measure protocol token distribution*. arXiv.org. Retrieved March 22, 2023, from https://arxiv.org/abs/2012.09306

Nakagawa, K., & Sakemoto, R. (2022). Cryptocurrency network factors and gold. *Finance Research Letters*, *46*, 102375.

Nakamoto, S. (2008). *Bitcoin: A peer-to-peer electronic cash system.* https://bitcoin.org/bitcoin.pdf.

Qin, K., Zhou, L., Afonin, Y., Lazzaretti, L., & Gervais, A. (2021). *CeFi vs. DeFi -- Comparing Centralized to Decentralized Finance*. arXiv.org. Retrieved 23 February 2022, from https://arxiv.org/abs/2106.08157.

Sai, A., Buckley, J., Fitzgerald, B., & Gear, A. (2021). Taxonomy of centralization in public blockchain systems: A systematic literature review. *Information Processing and Management*, *58*(4), 102584.

Taleb, N. (2015). How to (Not) Estimate Gini Coefficients for Fat Tailed Variables. arXiv.org. Retrieved 15 April 2022, from https://arxiv.org/abs/1510.04841.

Tsoukalas, G., & Falk, B. (2020). Token-Weighted Crowdsourcing. *Management Science*, *66*(9), 3843-3859.

Wang, G., Ma, X., & Wu, H. (2020). Are stablecoins truly diversifiers, hedges, or safe havens against traditional cryptocurrencies as their name suggests?. *Research In International Business And Finance*, *54*, 101225.

World Economic Forum. (2021). Decentralized Finance (DeFi) Policy-Maker Toolkit [White paper]. https://www.weforum.org/whitepapers/decentralized-finance-defi-policy-maker-toolkit

Wood, G. (2014). Ethereum: A Secure Decentralised Generalised Transaction Ledger. https://ethereum.github.io/yellowpaper/paper.pdf

Yermack, D. (2017). Corporate Governance and Blockchains. *Review Of Finance*, 21(1), 7-31.

Zhang, Z., Ren, D., Lan, Y., & Yang, S. (2022). Price competition and blockchain adoption in retailing markets. *European Journal Of Operational Research*, *300*(2), 647-660.
28

# Appendix

## A.1 Description of utilized factors

The following tables summarize the factors used in our univariate regressions.

### Table A.1: Financial factors for MKR and DAI

| | Description |
|---|---|
| **Price** | Price by day (USD) |
| **Return** | Daily return |
| **MktC** | Price (in USD) of the tokens times the circulating supply |
| **Volume** | Total amount (in tokens) of tokens transferred on Ethereum blockchain within a day |
| **Volume_usd** | Total amount (in USD) of tokens transferred on Ethereum blockchain within a day |
| **Volume_l** | Aggregated daily volume, measured in tokens from onchain transactions where each transaction was greater than $100,000 |
| **Volume_l_usd** | Aggregated daily volume, measured in USD from onchain transactions where each transaction was greater than $100,000 |

Note: The factors are provided by *intotheblock.com*.

### Table A.2: Transaction factors for MKR and DAI

| | Description |
|---|---|
| **AvgSize** | Total value of transactions (in tokens) divided by the number of transactions |
| **AvgSize_usd** | Total value of transactions (in USD) divided by the number of transactions |
| **Txncnt** | The number of valid transactions of tokens with in a day |
| **Volume_ex** | Sum of the amount (in tokens) entering an exchange plus the amount (in tokens) leaving an exchange. |
| **Volume_ex_usd** | Sum of the amount (in USD) entering an exchange plus the amount (in USD) leaving an exchange. |
| **Volume_dex** | Sum of the amount (in tokens) traded on Decentralized Exchanges (DEXes). |
| **Volume_dex_usd** | Sum of the amount (in USD) traded on Decentralized Exchanges (DEXes). |

Note: *Volume_dex* and *volume_dex_usd* are queried on *dune.xyz*. Other factors are provided by *intotheblock.com*.

### Table A.3: Network factors for MKR and DAI

| | Description |
|---|---|
| **TotalWithBlc** | The number of addresses that actually have a balance |
| **New** | The number of new addresses created daily |
| **Active** | The number of addresses that made a transaction |
| **Active ratio** | The percentage of addresses with a balance of tokens that had a transaction during a given period (Active Addresses / Addresses with a Balance). |

Note: The factors are provided by *intotheblock.com*.

### Table A.4: Twitter sentiment factors for MKR and DAI

| | Description |
|---|---|
| **Positive** | The number of Tweets that the texts used in the Tweets related to a given token have a positive connotation. |
| **Neutral** | The number of Tweets that the texts used in the Tweets related to a given token have a neutral connotation. |
| **Negative** | The number of Tweets that the texts used in the Tweets related to a given token have a negative connotation. |

Note: The twitter sentiment factors utilize machine learning algorithm to determine if the texts used in the Tweets related to a given token have a positive, neutral, or negative connotation. The factors are computed and provided by *intotheblock.com*.

### Table A.5: Collateral ratios

| | Description |
|---|---|
| **ETH_ratio** | The value (in USD) of ETH locked as collateral divided by the total value (in USD) of locked collateral in Maker protocol |

| | | |
|---|---|---|
| Stablecoin_ratio | The value (in USD) of stablecoins locked as collateral divided by the total value (in USD) of locked collateral in Maker protocol | |
| WBTC_ratio | The value (in USD) of Wrapped Bitcoin (WBTC) locked as collateral divided by the total value (in USD) of locked collateral in Maker protocol | |

Note: We focus on three types of collateral assets, including ETH, stablecoins and Wrapped Bitcoin (WBTC). The variables are queried on *dune.xyz*.

## A.2 Results for Granger test

The following tables summarize the results for Granger test. For the measurements based on governance polls, the number of observations is not enough for implementing Granger test. Therefore, we fill the null values using two different methods: linear interpolation and backward interpolation. We will discard the empirical findings that have problems of reserve causality

### Table A.6: Granger test results for MKR (linear interpolation)

**Panel A: Transaction**

| Null Hypothesis | Obs. | df | F-stat. | Prob. | Null Hypothesis | F-stat. | Prob. |
|---|---|---|---|---|---|---|---|
| Voters does not Granger Cause TxnCnt | 787 | 6 | 1.55 | 0.16 | **TxnCnt does not Granger Cause Voters** | **5.63** | **0.00** |
| Voters does not Granger Cause Volume_ex_usd | 803 | 7 | 1.53 | 0.16 | Volume_ex_usd does not Granger Cause Voters | 1.04 | 0.40 |

**PANEL B: Network**

| Null Hypothesis | Obs. | df | F-stat. | Prob. | Null Hypothesis | F-stat. | Prob. |
|---|---|---|---|---|---|---|---|
| Voters does not Granger Cause TotalWithBlc | 804 | 6 | 0.74 | 0.62 | **TotalWithBlc does not Granger Cause Voters** | **4.12** | **0.00** |
| Gini does not Granger Cause TotalWithBlc | 801 | 9 | 0.55 | 0.84 | TotalWithBlc does not Granger Cause Gini | 0.26 | 0.99 |
| Voters does not Granger Cause New | 804 | 6 | 2.33 | 0.03 | New does not Granger Cause Voters | 7.06 | 0.00 |
| Voters does not Granger Cause Active | 804 | 6 | 1.39 | 0.22 | **Active does not Granger Cause Voters** | **6.35** | **0.00** |

Note: This table reports the results for Granger tests based on Vector Autoregression (VAR) models. Column 'df' shows the optimal lag order. Using the optimal lag order, we run Granger tests for the hypothesizes stemming from our empirical findings. For each test, both F-statistics and probability are presented.

### Table A.7: Granger test results for MKR (backward interpolation)

**Panel A: Transaction**

| Null Hypothesis | Obs. | df | F-stat. | Prob. | Null Hypothesis | F-stat. | Prob. |
|---|---|---|---|---|---|---|---|
| Voters does not Granger Cause TxnCnt | 769 | 15 | 1.92 | 0.02 | **TxnCnt does not Granger Cause Voters** | **4.39** | **0.00** |
| Voters does not Granger Cause Volume_ex_usd | 799 | 11 | 1.01 | 0.43 | Volume_ex_usd does not Granger Cause Voters | 1.54 | 0.11 |

**PANEL B: Network**

| Null Hypothesis | Obs. | df | F-stat. | Prob. | Null Hypothesis | F-stat. | Prob. |
|---|---|---|---|---|---|---|---|
| Voters does not Granger Cause TotalWithBlc | 801 | 9 | 0.35 | 0.96 | **TotalWithBlc does not Granger Cause Voters** | **2.94** | **0.00** |
| Voters does not Granger Cause New | 790 | 20 | 1.59 | 0.05 | New does not Granger Cause Voters | 4.78 | 0.00 |
| Voters does not Granger Cause Active | 795 | 15 | 1.47 | 0.11 | **Active does not Granger Cause Voters** | **4.50** | **0.00** |
| Gini does not Granger Cause TotalWithBlc | 801 | 9 | 1.95 | 0.04 | TotalWithBlc does not Granger Cause Gini | 0.56 | 0.83 |

Note: This table reports the results for Granger tests based on Vector Autoregression (VAR) models. Column 'df' shows the optimal lag order. Using the optimal lag order, we run Granger tests for the hypothesizes stemming from our empirical findings. For each test, both F-statistics and probability are presented.

### Table A.8: Granger test results for MKR (measurements related to MKR distribution)

**PANEL A: Financial factors**

| Null Hypothesis | Obs. | df | F-stat. | Prob. | Null Hypothesis | F-stat. | Prob. |
|---|---|---|---|---|---|---|---|
| 10k-100k does not Granger Cause Volume | 804 | 6 | 1.70 | 0.12 | Volume does not Granger Cause 10k-100k | 1.19 | 0.31 |
| 10k-100k does not Granger Cause Volume_usd | 799 | 11 | 0.67 | 0.77 | Volume_usd does not Granger Cause 10k-100k | 0.44 | 0.94 |
| 10k-100k does not Granger Cause Volume_l | 804 | 6 | 2.02 | 0.06 | Volume_l does not Granger Cause 10k-100k | 1.47 | 0.19 |
| 10k-100k does not Granger Cause Volume_l_usd | 799 | 11 | 0.65 | 0.79 | Volume_l_usd does not Granger Cause 10k-100k | 0.44 | 0.94 |
| >100k does not Granger Cause Volume | 804 | 6 | 1.22 | 0.29 | Volume does not Granger Cause >100k | 0.82 | 0.55 |
| >100k does not Granger Cause Volume_usd | 804 | 6 | 2.55 | 0.02 | Volume_usd does not Granger Cause >100k | 0.07 | 1.00 |
| >100k does not Granger Cause Volume_l | 804 | 6 | 1.74 | 0.11 | Volume_l does not Granger Cause >100k | 0.89 | 0.50 |
| >100k does not Granger Cause Volume_l_usd | 799 | 11 | 0.83 | 0.61 | Volume_l_usd does not Granger Cause >100k | 0.10 | 1.00 |
| Delegate does not Granger Cause Volume_usd | 319 | 1 | 2.31 | 0.13 | **Volume_usd does not Granger Cause Delegate** | **2.81** | **0.09** |
| Delegate does not Granger Cause Volume_l_usd | 319 | 1 | 3.28 | 0.07 | Volume_l_usd does not Granger Cause Delegate | 0.59 | 0.44 |

| Panel B: Transaction | | | | | | | |
|---|---|---|---|---|---|---|---|
| Null Hypothesis | Obs. | df | F-stat. | Prob. | Null Hypothesis | F-stat. | Prob. |
| 10k-100k does not Granger Cause AvgSize_usd | 804 | 6 | 1.69 | 0.12 | AvgSize_usd does not Granger Cause 10k-100k | 0.78 | 0.58 |
| 10k-100k does not Granger Cause TxnCnt | 785 | 7 | 0.83 | 0.57 | **TxnCnt does not Granger Cause 10k-100k** | **8.61** | **0.01** |
| 10k-100k does not Granger Cause Volume_ex | 804 | 4 | 2.22 | 0.06 | Volume_ex does not Granger Cause 10k-100k | 0.48 | 0.75 |
| 10k-100k does not Granger Cause Volume_ex_usd | 799 | 11 | 0.88 | 0.56 | Volume_ex_usd does not Granger Cause 10k-100k | 0.58 | 0.84 |
| 10k-100k does not Granger Cause Volume_dex | 806 | 4 | 3.43 | 0.01 | Volume_dex does not Granger Cause 10k-100k | 0.15 | 0.96 |
| 10k-100k does not Granger Cause Volume_dex_usd | 801 | 9 | 1.08 | 0.37 | Volume_dex_usd does not Granger Cause 10k-100k | 0.25 | 0.99 |
| >100k does not Granger Cause AvgSizeMkr | 803 | 7 | 0.55 | 0.80 | **AvgSizeMkr does not Granger Cause >100k** | **4.37** | **0.00** |
| >100k does not Granger Cause AvgSize_usd | 804 | 6 | 3.72 | 0.00 | AvgSize_usd does not Granger Cause >100k | 0.27 | 0.95 |
| >100k does not Granger Cause TxnCnt | 785 | 7 | 0.61 | 0.75 | **TxnCnt does not Granger Cause >100k** | **5.20** | **0.03** |
| >100k does not Granger Cause Volume_ex | 806 | 4 | 3.52 | 0.01 | Volume_ex does not Granger Cause >100k | 0.41 | 0.80 |
| >100k does not Granger Cause Volume_ex_usd | 799 | 11 | 0.98 | 0.46 | Volume_ex_usd does not Granger Cause >100k | 0.28 | 0.99 |
| >100k does not Granger Cause Volume_dex | 806 | 4 | 3.45 | 0.01 | Volume_dex does not Granger Cause >100k | 0.60 | 0.66 |
| >100k does not Granger Cause Volume_dex_usd | 801 | 9 | 1.18 | 0.30 | Volume_dex_usd does not Granger Cause >100k | 0.36 | 0.95 |
| Delegate does not Granger Cause Volume_ex_usd | 318 | 2 | 7.10 | 0.00 | Volume_ex_usd does not Granger Cause Delegate | 0.74 | 0.48 |
| Delegate does not Granger Cause Volume_dex_usd | 319 | 1 | 5.08 | 0.02 | Volume_dex_usd does not Granger Cause Delegate | 0.05 | 0.82 |
| **PANEL C: Network** | | | | | | | |
| Null Hypothesis | Obs. | df | F-stat. | Prob. | Null Hypothesis | F-stat. | Prob. |
| 10k-100k does not Granger Cause TotalWithBlc | 801 | 9 | 2.33 | 0.01 | **TotalWithBlc does not Granger Cause 10k-100k** | **5.58** | **0.00** |
| 10k-100k does not Granger Cause New | 801 | 9 | 4.02 | 0.00 | New does not Granger Cause 10k-100k | 1.05 | 0.40 |
| 10k-100k does not Granger Cause Active | 803 | 7 | 1.86 | 0.07 | **Active does not Granger Cause 10k-100k** | **3.97** | **0.00** |
| >100k does not Granger Cause TotalWithBlc | 801 | 9 | 2.21 | 0.02 | **TotalWithBlc does not Granger Cause >100k** | **5.96** | **0.00** |
| >100k does not Granger Cause New | 804 | 6 | 4.39 | 0.00 | New does not Granger Cause >100k | 0.70 | 0.65 |
| >100k does not Granger Cause Active | 802 | 8 | 1.89 | 0.06 | **Active does not Granger Cause >100k** | **4.21** | **0.00** |
| Delegate does not Granger Cause TotalWithBlc | 312 | 8 | 8.77 | 0.00 | TotalWithBlc does not Granger Cause Delegate | 0.92 | 0.50 |
| Delegate does not Granger Cause New | 317 | 3 | 1.96 | 0.12 | New does not Granger Cause Delegate | 0.36 | 0.78 |
| Delegate does not Granger Cause Active | 314 | 6 | 3.59 | 0.00 | Active does not Granger Cause Delegate | 0.76 | 0.60 |
| **PANEL D: Twitter sentiment** | | | | | | | |
| Null Hypothesis | Obs. | df | F-stat. | Prob. | Null Hypothesis | F-stat. | Prob. |
| 10k-100k does not Granger Cause Positive | 771 | 7 | 0.67 | 0.70 | **Positive does not Granger Cause 10k-100k** | **2.52** | **0.01** |
| >100k does not Granger Cause Positive | 735 | 19 | 1.64 | 0.04 | **Positive does not Granger Cause >100k** | **2.27** | **0.00** |
| >100k does not Granger Cause Neutral | 750 | 14 | 3.45 | 0.00 | Neutral does not Granger Cause >100k | 0.64 | 0.83 |
| >100k does not Granger Cause Negative | 771 | 7 | 0.82 | 0.57 | Negative does not Granger Cause >100k | 0.64 | 0.73 |
| Delegate does not Granger Cause Positive | 294 | 3 | 1.71 | 0.16 | **Positive does not Granger Cause Delegate** | **6.51** | **0.00** |
| Delegate does not Granger Cause Negative | 304 | 1 | 10.05 | 0.00 | Negative does not Granger Cause Delegate | 1.21 | 0.27 |

Note: This table reports the results for Granger tests based on Vector Autoregression (VAR) models. Column 'df' shows the optimal lag order. Using the optimal lag order, we run Granger tests for the hypothesizes stemming from our empirical findings. For each test, both F-statistics and probability are presented.

Table A.9: Granger test results for DAI (linear interpolation)

| PANEL A: Financial factors | | | | | | | |
|---|---|---|---|---|---|---|---|
| Null Hypothesis | Obs. | df | F-stat. | Prob. | Null Hypothesis | F-stat. | Prob. |
| Voters does not Granger Cause Price | 789 | 18 | 2.60 | 0.00 | Price does not Granger Cause Voters | 1.23 | 0.23 |
| Voters does not Granger Cause Volume | 703 | 7 | 0.20 | 0.99 | Volume does not Granger Cause Voters | 1.64 | 0.12 |
| Voters does not Granger Cause Volume_usd | 690 | 20 | 8.05 | 0.00 | **Volume_usd does not Granger Cause Voters** | **1.52** | **0.07** |
| Voters does not Granger Cause Volume_l | 704 | 6 | 0.24 | 0.97 | Volume_l does not Granger Cause Voters | 1.63 | 0.14 |
| Voters does not Granger Cause Volume_l_usd | 704 | 6 | 0.24 | 0.96 | Volume_l_usd does not Granger Cause Voters | 1.61 | 0.14 |
| **PANEL B: Network** | | | | | | | |
| Null Hypothesis | Obs. | df | F-stat. | Prob. | Null Hypothesis | F-stat. | Prob. |
| Voters does not Granger Cause ΔNew | 704 | 6 | 1.07 | 0.38 | ΔNew does not Granger Cause Voters | 0.98 | 0.44 |
| Gini does not Granger Cause ΔActiveRatio | 695 | 15 | 0.23 | 1.00 | ΔActiveRatio does not Granger Cause Gini | 0.36 | 0.99 |
| **PANEL C: Twitter sentiment** | | | | | | | |
| Null Hypothesis | Obs. | df | F-stat. | Prob. | Null Hypothesis | F-stat. | Prob. |
| Voters does not Granger Cause Positive | 763 | 6 | 0.40 | 0.88 | **Positive does not Granger Cause Voters** | **1.85** | **0.09** |

Note: This table reports the results for Granger tests based on Vector Autoregression (VAR) models. Column 'df' shows the optimal lag order. Using the optimal lag order, we run Granger tests for the hypothesizes stemming from our empirical findings. For each test, both F-statistics and probability are presented.

### Table A.10: Granger test results for DAI (backward interpolation)

| PANEL A: Financial factors | | | | | | | |
|---|---|---|---|---|---|---|---|
| Null Hypothesis | Obs. | df | F-stat. | Prob. | Null Hypothesis | F-stat. | Prob. |
| Voters does not Granger Cause Price | 792 | 15 | 0.84 | 0.63 | Price does not Granger Cause Voters | 0.47 | 0.96 |
| Voters does not Granger Cause Volume | 693 | 17 | 4.55 | 0.00 | **Volume does not Granger Cause Voters** | **5.36** | **0.00** |
| Voters does not Granger Cause Volume_usd | 690 | 20 | 9.09 | 0.00 | **Volume_usd does not Granger Cause Voters** | **2.16** | **0.00** |
| Voters does not Granger Cause Volume_l | 693 | 17 | 4.59 | 0.00 | **Volume_l does not Granger Cause Voters** | **5.37** | **0.00** |
| Voters does not Granger Cause Volume_l_usd | 693 | 17 | 4.56 | 0.00 | **Volume_l_usd does not Granger Cause Voters** | **5.32** | **0.00** |
| **PANEL B: Network** | | | | | | | |
| Null Hypothesis | Obs. | df | F-stat. | Prob. | Null Hypothesis | F-stat. | Prob. |
| Voters does not Granger Cause ΔNew | 701 | 9 | 1.38 | 0.19 | ΔNew does not Granger Cause Voters | 0.55 | 0.84 |
| Gini does not Granger Cause ΔActiveRatio | 690 | 20 | 0.97 | 0.50 | **ΔActiveRatio does not Granger Cause Gini** | **2.94** | **0.00** |
| **PANEL C: Twitter sentiment** | | | | | | | |
| Null Hypothesis | Obs. | df | F-stat. | Prob. | Null Hypothesis | F-stat. | Prob. |
| Voters does not Granger Cause Positive | 698 | 19 | 1.20 | 0.25 | **Positive does not Granger Cause Voters** | **3.84** | **0.00** |

### Table A.11: Granger test results for DAI (measurements related to MKR distribution)

| PANEL A: Financial factors | | | | | | | |
|---|---|---|---|---|---|---|---|
| Null Hypothesis | Obs. | df | F-stat. | Prob. | Null Hypothesis | F-stat. | Prob. |
| 10k-100k does not Granger Cause Price | 788 | 19 | 1.31 | 0.17 | **Price does not Granger Cause 10k-100k** | **4.01** | **0.00** |
| 10k-100k does not Granger Cause ΔMktC | 706 | 4 | 3.97 | 0.00 | ΔMktC does not Granger Cause 10k-100k | 0.22 | 0.93 |
| 10k-100k does not Granger Cause Volume | 706 | 4 | 3.96 | 0.00 | Volume does not Granger Cause 10k-100k | 0.22 | 0.93 |
| 10k-100k does not Granger Cause Volume_usd | 706 | 4 | 3.89 | 0.00 | Volume_usd does not Granger Cause 10k-100k | 0.24 | 0.92 |
| 10k-100k does not Granger Cause Volume_l | 706 | 4 | 3.74 | 0.01 | Volume_l does not Granger Cause 10k-100k | 0.22 | 0.93 |
| 10k-100k does not Granger Cause Volume_l_usd | 706 | 4 | 3.71 | 0.01 | Volume_l_usd does not Granger Cause 10k-100k | 0.22 | 0.93 |
| >100k does not Granger Cause Price | 788 | 19 | 0.75 | 0.77 | **Price does not Granger Cause >100k** | **4.32** | **0.00** |
| >100k does not Granger Cause ΔMktC | 706 | 4 | 4.60 | 0.00 | ΔMktC does not Granger Cause >100k | 0.65 | 0.63 |
| >100k does not Granger Cause Volume | 709 | 1 | 65.58 | 0.00 | Volume does not Granger Cause >100k | 0.00 | 1.00 |
| >100k does not Granger Cause Volume_usd | 709 | 1 | 69.09 | 0.00 | Volume_usd does not Granger Cause >100k | 0.14 | 0.71 |
| >100k does not Granger Cause Volume_l | 709 | 1 | 60.72 | 0.00 | Volume_l does not Granger Cause >100k | 0.00 | 0.99 |
| >100k does not Granger Cause Volume_l_usd | 709 | 1 | 59.98 | 0.00 | Volume_l_usd does not Granger Cause >100k | 0.00 | 0.99 |
| **Panel B: Transaction** | | | | | | | |
| Null Hypothesis | Obs. | df | F-stat. | Prob. | Null Hypothesis | F-stat. | Prob. |
| 10k-100k does not Granger Cause AvgSizeDai | 709 | 1 | 23.38 | 0.00 | AvgSizeDai does not Granger Cause 10k-100k | 0.06 | 0.81 |
| 10k-100k does not Granger Cause AvgSize_usd | 709 | 1 | 23.03 | 0.00 | AvgSize_usd does not Granger Cause 10k-100k | 0.06 | 0.80 |
| 10k-100k does not Granger Cause TxnCnt | 690 | 6 | 2.48 | 0.02 | TxnCnt does not Granger Cause 10k-100k | 1.66 | 0.13 |
| >100k does not Granger Cause AvgSizeDai | 709 | 1 | 67.48 | 0.00 | AvgSizeDai does not Granger Cause >100k | 0.02 | 0.90 |
| >100k does not Granger Cause AvgSize_usd | 709 | 1 | 66.68 | 0.00 | AvgSize_usd does not Granger Cause >100k | 0.02 | 0.90 |
| >100k does not Granger Cause TxnCnt | 686 | 8 | 2.46 | 0.01 | TxnCnt does not Granger Cause >100k | 1.53 | 0.14 |
| >100k does not Granger Cause Volume_ex | 704 | 6 | 3.33 | 0.00 | **Volume_ex does not Granger Cause >100k** | **4.16** | **0.00** |
| >100k does not Granger Cause Volume_ex_usd | 704 | 6 | 3.31 | 0.00 | **Volume_ex_usd does not Granger Cause >100k** | **4.14** | **0.00** |
| Delegate does not Granger Cause TxnCnt | 308 | 2 | 0.39 | 0.68 | TxnCnt does not Granger Cause Delegate | 0.44 | 0.65 |
| Delegate does not Granger Cause Volume_ex | 314 | 6 | 0.22 | 0.97 | Volume_ex does not Granger Cause Delegate | 0.55 | 0.77 |
| Delegate does not Granger Cause Volume_ex_usd | 314 | 6 | 0.22 | 0.97 | Volume_ex_usd does not Granger Cause Delegate | 0.55 | 0.77 |
| Delegate does not Granger Cause ΔVolume_dex | 315 | 5 | 0.57 | 0.72 | ΔVolume_dex does not Granger Cause Delegate | 0.53 | 0.75 |
| Delegate does not Granger Cause ΔVolume_dex_usd | 315 | 5 | 0.57 | 0.72 | ΔVolume_dex_usd does not Granger Cause Delegate | 0.53 | 0.75 |
| **PANEL C: Twitter sentiment** | | | | | | | |
| Null Hypothesis | Obs. | df | F-stat. | Prob. | Null Hypothesis | F-stat. | Prob. |
| 10k-100k does not Granger Cause Positive | 748 | 9 | 5.75 | 0.00 | Positive does not Granger Cause 10-100k | 0.72 | 0.69 |
| >100k does not Granger Cause Positive | 748 | 9 | 8.67 | 0.00 | Positive does not Granger Cause >100k | 0.41 | 0.93 |

| Null Hypothesis | Obs. | df | F-stat. | Prob. | Null Hypothesis | F-stat. | Prob. |
|---|---|---|---|---|---|---|---|
| >100k does not Granger Cause Neutral | 778 | 3 | 0.49 | 0.69 | Neutral does not Granger Cause >100k | 0.14 | 0.94 |
| >100k does not Granger Cause Negative | 698 | 19 | 2.14 | 0.00 | Negative does not Granger Cause >100k | 1.65 | 0.04 |
| Delegate does not Granger Cause Positive | 303 | 1 | 6.05 | 0.01 | **Positive does not Granger Cause Delegate** | **16.52** | **0.00** |
| Delegate does not Granger Cause Negative | 299 | 2 | 7.58 | 0.00 | Negative does not Granger Cause Delegate | 1.56 | 0.21 |

Note: This table reports the results for Granger tests based on Vector Autoregression (VAR) models. Column 'df' shows the optimal lag order. Using the optimal lag order, we run Granger tests for the hypothesizes stemming from our empirical findings. For each test, both F-statistics and probability are presented.

**Table A.12: Granger test results for collateral ratios**

| Null Hypothesis | Obs. | df | F-stat. | Prob. | Null Hypothesis | F-stat. | Prob. |
|---|---|---|---|---|---|---|---|
| 10k-100k does not Granger Cause $\Delta$ETH_ratio | 700 | 4 | 2.80 | 0.03 | $\Delta$ETH_ratio does not Granger Cause 10-100k | 0.74 | 0.57 |
| 10k-100k does not Granger Cause $\Delta$Stablecoin_ratio | 700 | 4 | 1.95 | 0.10 | $\Delta$Stablecoin_ratio does not Granger Cause 10-100k | 1.53 | 0.19 |
| 10k-100k does not Granger Cause $\Delta$WBTC_ratio | 684 | 20 | 0.79 | 0.73 | **$\Delta$WBTC_ratio does not Granger Cause 10-100k** | **4.31** | **0.00** |
| >100k does not Granger Cause $\Delta$ETH_ratio | 698 | 6 | 3.52 | 0.00 | $\Delta$ETH_ratio does not Granger Cause >100k | 1.09 | 0.36 |
| >100k does not Granger Cause $\Delta$Stablecoin_ratio | 698 | 6 | 2.79 | 0.01 | **$\Delta$Stablecoin_ratio does not Granger Cause >100k** | **1.90** | **0.08** |
| >100k does not Granger Cause $\Delta$WBTC_ratio | 685 | 19 | 0.91 | 0.57 | **$\Delta$WBTC_ratio does not Granger Cause >100k** | **4.07** | **0.00** |
| Delegate does not Granger Cause $\Delta$ETH_ratio | 319 | 1 | 4.94 | 0.03 | $\Delta$ETH_ratio does not Granger Cause Delegate | 0.65 | 0.42 |
| Delegate does not Granger Cause $\Delta$Stablecoin_ratio | 319 | 1 | 4.31 | 0.04 | $\Delta$Stablecoin_ratio does not Granger Cause Delegate | 1.12 | 0.29 |